\documentclass[aps,pra,preprint,showpacs,preprintnumbers,amsmath,amssymb,floatfix,a4paper,longbibliography,superscriptaddress]{revtex4-2}
\usepackage{amsmath}
\usepackage{amsfonts}
\usepackage{amssymb}
\usepackage{bm}
\usepackage{bbm}
\usepackage{color}
\usepackage{graphicx}
\usepackage{hyperref}
\usepackage{dcolumn}

\newcommand{\figwidth}{0.5\paperwidth}
\newcommand{\colw}{0.1\columnwidth}
\newcommand{\Efield}{F}
\newcommand{\EBS}{\Efield_{\rm BS}}
\newcommand{\Fred}{\widetilde{F}}
\newcommand{\W}{\mathit{\Gamma}}
\newcommand{\WTong}{\W_{\rm fit}}
\newcommand{\WPPT}{\W_{\rm PPT}}
\newcommand{\WH}{\W_{\rm H}}
\newcommand{\WSAE}{\W_{\rm SAE}}
\newcommand{\Wpert}{\W_{\rm pert}}
\newcommand{\WFC}{\W_{{\rm FC}}}
\newcommand{\afit}{{\rm \textsl a}}
\newcommand{\rmd}{\mathrm{d}}
\DeclareMathOperator{\imag}{Im}

\begin{document}
\title{Static-field ionization model of He-like ions for diagnostics of light field intensity}
\date{\today}
\author{Erik L\"otstedt}
\email{lotstedt@chem.s.u-tokyo.ac.jp}
\affiliation{Department of Chemistry, School of Science, The University of Tokyo,
7-3-1 Hongo, Bunkyo-ku, Tokyo 113-0033, Japan}
\author{Marcelo F. Ciappina}
\affiliation{Institute of Physics of the ASCR, ELI-Beamlines project, Na Slovance 2, 182 21 Prague, Czech Republic}
\affiliation{ICFO -- Institut de Ciencies Fotoniques, The Barcelona Institute of Science and Technology, 08860 Castelldefels (Barcelona)}
\author{Kaoru Yamanouchi}
\affiliation{Department of Chemistry, School of Science, The University of Tokyo,
7-3-1 Hongo, Bunkyo-ku, Tokyo 113-0033, Japan}

\begin{abstract}
We study static-field  ionization of He-like ions with nuclear charge number in the range of
$2\le Z \le 36$. Both the tunneling and over-the-barrier regimes are considered. 
We calculate the ionization rates  by three approximate methods: a 
fitting formula based on the 
Perelemov-Popov-Terent'ev 
(PPT) 
formula,
 a perturbative expansion in powers of $1/Z$, and a single-active electron approximation, and compare them with reference   ionization rates computed 
by
 the 
multiconfiguration time-dependent Hartree-Fock (MCTDHF) approach. 
The relative deviation of the rates 
computed 
 by the PPT-based fitting formula and 
 the third-order perturbation theory 
  from the 
  rates computed by the MCTDHF approach 
   is found to be around 10\%.
   We  discuss
    quantitatively the contribution from
  the exchange interaction 
during 
the 
ionization by comparing the single-active electron rates and the reference 
rates  in
 which multielectron effects are included.
We find that a single-active electron approximation, 
where the exchange interaction is 
neglected,
 results in 
the overestimation of the ionization rates 
by  30\% for He and 2\% for He-like Kr, showing that the magnitude of the effect of 
the exchange interaction scales approximately as $1/Z$.
\end{abstract}

\maketitle
%

%

\section{Introduction}\label{Intro}
Strong-field ionization of atoms has been proposed as a means 
of estimating
 the peak intensity of  laser pulses having an intensity of $I>10^{20}$ W/cm$^2$ \cite{Hetzheim2009,BaukeHetzheimetal2011,Ciappinaetal2019,CiappinaPopruzhenko_2020,CiappinaBulanovetal2020}. 
Such ultra-intense laser pulses are becoming available at several facilities worldwide, two examples being the ELI Beamlines \cite{ELIbeamlines2015} 
and the Shanghai super-intense ultrafast laser facility \cite{Guo2018}.
In \cite{Ciappinaetal2019,CiappinaPopruzhenko_2020}, it was shown that the relative yields of different ionic charge states are highly sensitive to the peak intensity of the laser 
pulse and that the final charge state distribution 
 can be used to estimate the laser 
field
 intensity at 
the 
beam focus. Moreover, it was concluded in \cite{CiappinaPopruzhenko_2020} 
that the highest accuracy can be achieved in the determination of the laser 
field intensity when the ionization yields of He- and H-like ions are adopted.

In order to 
relate quantitatively
 the measured ion yield 
 to
  the laser field intensity, it is necessary to simulate the ion yields using 
  reliable
   formulas for the tunneling and 
   over-the-barrier (OTB) 
   static-field
    ionization rates. 
 Ionization rates  
for H-like ions are well established through the 
tunneling-rate formula originally derived by Smirnov and Chibisov \cite{SMIRNOVCHIBISOV1966}, 
which has also been referred to as the 
 Perelemov-Popov-Terent'ev (PPT) formula   \cite{PPT_1966,Popov_2004,V_Popruzhenko_2014} or the Ammosov-Delone-Krainov (ADK) \cite{Ammosov1986} formula,
providing
 reliable estimates of the ionization rates in the tunneling regime.
 Within the single-active electron approximation,
the ADK theory has been extended to molecular systems 
\cite{TongZhaoLin2002,Kjeldsen_2004,Zhao2011,KornevZon2015,KornevChernovZon2017}. 
Attempts have also been made to include many-electron effects in static-field tunneling ionization
\cite{FisherMaronPitaevskii1998,Brabecetal2005}.
When the field intensity is
sufficiently weak, the weak-field asymptotic theory (WFAT)
\cite{TolstikhinMorishitaMadsen2011} provides a rigorous expansion of the pre-exponential
factor in the tunneling rate in powers of the field strength.
 The 
WFAT has been extended to many-electron systems \cite{TolstikhinMadsenMorishita2014,TolstikhinaMorishitaTolstikhin2014,Trinh_2015,Trinh_2016} and has been applied to molecular ionization \cite{YueBauchMadsen2017,DnestryanTolstikhinJensenMadsen2019}.
In \cite{Trinh_2016}, the WFAT  was applied to He-like ions for 
an arbitrary value of 
the nuclear charge number $Z$.

In order to estimate a static-field ionization rate
 in the  OTB regime, 
 we need to adopt numerical methods because
  there are no analytical formulas available.
Static-field ionization 
rates of 
H \cite{NicolaidesThemelis1992,Karlsson_1992,TrinhTolstikhinMadsenMorishita2013}, 
He \cite{NicolaidesThemelis1993,ScrinziGeisslerBrabec1999,Scrinzi2000} and 
H$_2$ \cite{Saenz2000} have been studied numerically by the complex scaling methods,
which have also been applied to  
 many-electron atoms \cite{Jagau2016} and 
 diatomic 
 and triatomic molecules (CO, N$_2$, O$_2$, CO$_2$) \cite{Majety_2015,MajetyScrinzi2015}. 
 In Ref.~\cite{ParkerArmstrongBocaTaylor2009}, static-field ionization rates of He were obtained by real-time integration of 
 the time-dependent Schr\"odinger equation (TDSE). 
 However, to the best of our knowledge, no numerical studies 
 have been reported so far on the tunneling and OTB ionization rates of
  He-like ions. 
  It has been awaited that reliable methods for estimating ionization rates for He-like ions valid in both the tunneling and the OTB regimes are established, so that the scheme for estimating the laser field intensity proposed in \cite{Ciappinaetal2019} will be realized. The advantage of treating He-like ions is that the effect of the electron-electron correlation in the static-field ionization processes can be varied systematically by the change in the nuclear charge number $Z$.

 In the present study, we
  investigate the static-field tunneling and OTB ionization of He-like ions with 
  the 
  nuclear charge number $Z$ in the range of $2\le Z\le 36$. We 
  compare
    (i) ionization rates obtained by
  the PPT formula, (ii) 
  ionization rates obtained by 
   a PPT-based fitting formula proposed by Tong and Lin  \cite{TongLin_2005}, 
 (iii) 
 ionization rates calculated by a perturbative expansion in powers of $1/Z$, (iv) ionization rates calculated by the single-active electron (SAE) approximation, and (v) reference 
 ionization
 rates obtained by 
 the  multiconfiguration time-dependent Hartree-Fock (MCTDHF) method \cite{ZanghelliniScrinzi2003,KatoKono2004}, in which 
 the electron-electron interaction is fully accounted for. 
We compare the  deviation of the 
approximate rates from the rates calculated by the MCTDHF method, and find that both the 
approach (ii) and 
the 
approach (iii) result in rates which deviate by about 10\% from the MCTDHF rates. 
Considering that the approach (ii) is easier 
to be applied than the approach (iii), we recommend the approach (ii) to be used in
 the evaluation of static-field ionization rates of He-like ions.

\section{Theory}\label{Sec:Theory}
In this section, we describe the theoretical methods 
that we employ in the present study.
Atomic units (a.u.) are used throughout the 
paper unless
 otherwise indicated. For the conversion from a.u.\ to SI units, we have $1\;{\rm a.u.} = 4.134\times 10^{16}\;{\rm s}^{-1}$ for ionization rates, and $1\;{\rm a.u.} = 5.142\times 10^{11}$ V/m for electric fields.

In 
 the calculation of the static-field ionization rates,
we employ the nonrelativistic   TDSE. Although we will present 
results for the ionization of He-like ions with 
the large charge number $Z$ up to $Z = 36$, where the intensity of a laser field
required for ionization reaches
 $10^{22}$ W/cm$^2$, it has been shown that the ionization process itself is essentially  nonrelativistic
for ions with $Z<20$ \cite{MilosevicKrainovBrabec2002,Ciappinaetal2019}. 
As long as the ionization potential $I_p$ of an atom is much smaller than the rest energy of an electron,
$I_p\ll c^2$, where $c\approx 137$ a.u., 
the relativistic 
effects in the ground state wave function can be neglected.
Using $I_p\approx Z^2/2$ for large $Z$, we see that $I_p\ll c^2$ is satisfied even for $Z=36$ because $I_p/c^2=36^2/(2c^2)\approx 0.03$.

The static-field ionization  rates obtained in this study can be used to  estimate the 
ionization probability in a low-frequency laser field 
as long as the static-field ionization rates can be  used as 
instantaneous ionization rates. 
This procedure is valid if the Keldysh 
parameter $\gamma=\sqrt{2I_p}\omega/\Efield$, where $\omega$ is the 
angular frequency of the laser field and $\Efield$ is the peak field strength, is  
smaller than 1 \cite{Keldysh1964,Ciappinaetal2019}. 
In this study, 
we consider values of $\Efield$ so that $\gamma<1$. 
For example,  for He ($Z=2$), we consider $\Efield\approx 0.4$ a.u., which 
gives $\gamma=0.2$ assuming $\omega=0.057$ a.u.\ (corresponding to a wavelength of 800 nm), and for He-like Kr ($Z=36$), we consider $\Efield\approx 2\times 10^3$ a.u., 
which gives 
$\gamma=0.001$ at $\omega=0.057$ a.u.

\subsection{$Z$-scaling}\label{subsec:Zscaling}
The TDSE for a He-like ion with 
the 
 charge number $Z$ in a static electric field $\bm{\Efield}=\Efield\hat{\bm{z}}$ reads
\begin{equation}
\label{TDSE_unscaled}
i\frac{\partial}{\partial t}\Psi(\bm{r}_1,\bm{r}_2,t)=
\left[H+\Efield(z_1+z_2)\right]
\Psi(\bm{r}_1,\bm{r}_2,t),
\end{equation}
where $\bm{r}_i=(x_i,y_i,z_i)$ is the coordinate of the $i$th electron, and  the field-free Hamiltonian is defined as
\begin{equation}
H=\sum_{i=1,2}\left(-\frac{1}{2}\frac{\partial^2}{\partial\bm{r}^2_i} -\frac{Z}{r_i} \right)
+\frac{1}{r_{12}},
\end{equation}
with $r_{12}=|\bm{r}_1-\bm{r}_2|$.
 After scaling  the spatial coordinates and time according to \cite{ScherrKnight1963}
\begin{equation}
\bm{r}_i=\frac{2\bm{r}'_i}{Z},\qquad t = \frac{4t'}{Z^2},
\end{equation}
and rewriting the TDSE \eqref{TDSE_unscaled} in the scaled coordinates, we obtain 
\begin{equation}
\label{TDSE_scaled}
i\frac{\partial}{\partial t'}\Psi'(\bm{r}'_1,\bm{r}'_2,t')=
\left[H'+\Efield'(z'_1+z'_2)\right]
\Psi'(\bm{r}'_1,\bm{r}'_2,t'),
\end{equation}
where 
\begin{equation}\label{scaled_H0}
H'=\sum_{i=1,2}\left(-\frac{1}{2}\frac{\partial^2}{\partial\bm{r}'^2_i} -\frac{2}{r'_i} \right)
+\frac{2}{Z}\frac{1}{r'_{12}},
\end{equation}
and $\Psi'(\bm{r}'_1,\bm{r}'_2,t')=(2/Z)^3\Psi(\bm{r}_1,\bm{r}_2,t)$. 
The scaling of the amplitude of the electric  field 
is
\begin{equation}\label{FieldScaling}
\Efield'=\frac{8}{Z^3}\Efield.
\end{equation}

The transformation described above implies that a He-like atomic ion with 
the 
charge number $Z$ is described by the same TDSE as He 
and that
 the electron-electron interaction is scaled by $2/Z$, 
 indicating that the relative magnitude  of the electron-electron interaction 
 compared to the electron-nucleus interaction decreases as $Z$ increases.

For 
the ionization rate, 
 we have the following scaling relation,
\begin{equation}\label{scalingrelation_rates}
\W(\Efield)=\frac{Z^2}{4}\W'(\Efield'),
\end{equation}
where $\W(\Efield)$ is the rate 
we obtain 
 by solving Eq.~\eqref{TDSE_unscaled} at the field 
 $\Efield$ and 
$\W'(\Efield')$ is the rate 
we obtain 
 by solving Eq.~\eqref{TDSE_scaled} at the field $\Efield'= 8 \Efield/Z^3$.

In the following part of the paper,  symbols with a prime 
 always refer to 
the quantities in 
the transformed system \eqref{TDSE_scaled} 
and symbols without a prime
  refer to 
those in 
the original system \eqref{TDSE_unscaled}.

\subsection{MCTDHF}\label{subsec:MCTDHF}
In order to solve the  scaled TDSE \eqref{TDSE_scaled} numerically, we employ the MCTDHF method 
\cite{ZanghelliniScrinzi2003,KatoKono2004,HochstuhlHinzBonitz2014,%
IshikawaSato_Review2015,Loetstedt2017,SatoOrimoetal2018}. In the MCTDHF method, the  two-electron  wave function is expanded in terms   of time-dependent Slater determinants 
$\|\phi'_i(t')\alpha\,\phi'_j(t')\beta\|$,
\begin{equation}\label{MCTDHF_def}
\Psi'(t')=\sum_{i,j=1}^M C'_{ij}(t')\|\phi'_i(t')\alpha\,\phi'_j(t')\beta\|,
\end{equation}
where  $\phi'_i(\bm{r}', t')$ is a time-dependent spatial orbital, $M$ is the total number of orbitals, $\alpha$ (spin up) and 
$\beta$ (spin down) are spin functions, and $C'_{ij}(t')$ is a time-dependent  configuration-interaction (CI) coefficient. The equations of motion for the spin-orbitals and the CI coefficients are derived by applying the time-dependent variational principle \cite{ZanghelliniScrinzi2003,KatoKono2004}. 
The MCTDHF method has previously been applied to He 
in the calculation of 
two-photon ionization probabilities \cite{Hochstuhl2011}, high-harmonic spectra \cite{SatoIshikawaetal2016}, above-threshold photoelectron spectra \cite{Orimoetal2018}, 
and strong-field excitation probabilities \cite{LotstedtSzidarovszkyetal2020}. 

It should be noted here 
 that the MCTDHF method is different from other commonly employed close-coupling approaches 
\cite{Zhang_1995,SMYTH1998,LaulanBachau2003,ParkerArmstrongBocaTaylor2009} in that products of 
single-electron orbitals are used 
in the expansion of 
 the time-dependent wave function, while  in \cite{Zhang_1995,SMYTH1998,LaulanBachau2003,ParkerArmstrongBocaTaylor2009} the two-electron wave function is expanded 
in terms of two-electron radial basis functions multiplied with two-electron spherical harmonics.

We solve the MCTDHF equations of motion in the same way as described in \cite{LotstedtKatoYamanouchi2017,LotstedtSzidarovszkyetal2020}. 
The details of the numerical implementation of the MCTDHF method are  
summarized in 
Appendix ~\ref{App_MCTDHF}.

In order to obtain the 
ionization 
 rate $\W$, we solve
the scaled TDSE \eqref{TDSE_scaled} starting from the ground state, $\Psi'(t'=0)=\Psi'_0$, and calculate the time-dependent population of the ground state, $p_0(t')=|\langle \Psi'_0|\Psi'(t)\rangle|^2$. 
For 
$t' < t'_0$ 
($t'_0\approx 50$ a.u.),
 $p_0(t')$ oscillates due to the transitions to 
the
excited states induced by the sudden turn-on of the 
static 
field. For $t'>t'_0$, $p_0(t')$ decreases exponentially
and can be fitted  to an exponential function $e^{-\W' t' -q'}$  with 
the 
constant  $\W'$, representing
 the scaled ionization rate, 
 and $q'$ \cite{Scrinzi2000}.
 After the simulations with
  different values of the 
  static 
  field $F'$, the scaled field-dependent ionization rate $\W'(\Efield')$ 
  is obtained, from which  the 
 ionization rate of a He-like ion is 
 calculated 
 by the scaling of Eq.~\eqref{scalingrelation_rates}.
 However, if the ionization rate is very large or very small, it becomes difficult to obtain the ionization rate by this approach because $p_0(t')$ cannot be fitted well to 
 $e^{-\W' t' -q'}$ when it changes too slowly or too rapidly.
Consequently, the scaled fields need to be limited practically in the range of $0.1\lesssim\Efield' \lesssim 1$ a.u.

\subsection{Perturbation theory}\label{subsec:PertTheory}
In the limit
of 
 $Z\to\infty$, the scaled Eq.~\eqref{TDSE_scaled} decouples and becomes an equation for two independent He$^+$ ions. Therefore, in the large  $Z$ limit,  the scaled ionization rate for a He-like ion can be expressed as
\begin{equation}\label{ExactScaling_WH}
\W'(\Efield')\underset{Z\to \infty}{=}8\WH(\Efield),
\end{equation}
where $\WH(\Efield)$ is the ionization rate of H evaluated at 
the 
unscaled field of $\Efield=\Efield'/8$.

For finite but large values of $Z$, we can treat the electron-electron interaction term $2/(Zr'_{12})$ in the scaled Hamiltonian \eqref{scaled_H0} as a perturbation, and  evaluate corrections 
using  perturbation theory. By inserting the expansions 
\begin{equation}
\Psi'(\bm{r}'_1,\bm{r}'_2)=\sum_{n=0}^\infty \frac{\chi'_n(\bm{r}'_1,\bm{r}'_2)}{Z^n}
\end{equation}
and
\begin{equation}
\epsilon'=\sum_{n=0}^\infty \frac{\epsilon'_n}{Z^n}
\end{equation}
into the 
eigen equation,
\begin{equation}
H'(e^{i\Theta}\bm{r}'_1,e^{i\Theta}\bm{r}'_2)\Psi'(\bm{r}'_1,\bm{r}'_2)=\epsilon' \Psi'(\bm{r}'_1,\bm{r}'_2),
\end{equation}
 where $H'(e^{i\Theta}\bm{r}'_1,e^{i\Theta}\bm{r}'_2)$ is the 
 scaled complex-rotated
  Hamiltonian  \eqref{scaled_H0}  \cite{Reinhardt1982}, we derive   equations relating $\chi'_n$ and $\epsilon'_n$ at different orders of $1/Z$.   Throughout our calculations we use 
the complex rotation parameter $\Theta =0.2$. The equations up to 
the third order 
 are given in Appendix~\ref{PertTheoryAppendix_subsec1}. 
 After the 
field-dependent correction terms $\epsilon'_n(\Efield')$ 
for
 the eigenvalue of the ground state 
are calculated,
 the unscaled ionization rate 
 corrected up to 
 the 
 order $N$ at a given value of $Z$ is given by 
 \begin{equation}\label{perttheory_W}
 \Wpert^N(\Efield') = \frac{Z^2}{4}\sum_{n=0}^N\frac{\lambda'_n(\Efield')}{Z^n},
 \end{equation}
 with
 \begin{equation}\label{pertCoeffs}
\lambda'_n(\Efield') =-2 \imag\left[\epsilon'_n(\Efield')\right].
 \end{equation}
 It should be noted here  that the correction terms $\epsilon'_n$ contain corrections both to the eigenenergy (the real part of $\epsilon'_n$) and to the ionization rate (the imaginary part of $\epsilon'_n$) of the ground state 
 and that the rate given by Eq.~\eqref{perttheory_W} contains contributions 
from all orders in the field $\Efield'$, because the interaction with the field is included in the zeroth-order Hamiltonian.

\subsection{Single-active electron approximation}\label{subsec:SAE}
In the single-active electron approximation (SAE), the SAE Hamiltonian reads
\begin{equation}\label{SAEHamiltonian}
H'_{\rm SAE}(\bm{r}')=-\frac{1}{2}\frac{\partial^2}{\partial\bm{r}'^2} +V(\bm{r}') +\Efield' z',
\end{equation}
where $V(\bm{r}')$ is an effective potential. 
We consider below the SAE approximation to the scaled TDSE \eqref{TDSE_scaled}.

A commonly used effective potential for He-like ions reads \cite{Scrinzi2000}
\begin{equation}\label{V1}
V_{1}(\bm{r}')=-\frac{2-\frac{2}{Z}+\frac{2}{Z}e^{-\zeta' r'}}{r'},
\end{equation}
where the parameter $\zeta'$ is adjusted so that the absolute value of the ground state energy 
becomes equal to the ionization potential of the corresponding two-electron system.
For He, we additionally consider a polarization potential in order to model the effect of the electric field on the inner electron,
that is, the electron left bound after the ejection of the other electron. The polarization potential is thus defined as
\begin{equation}\label{Vpol}
V_{\rm pol}(\bm{r}) = -\Efield \alpha_{{\rm He}^+} \cos\theta \frac{1-e^{-\eta r}\left(1+\eta r+\frac{1}{2}\eta^2 r^2\right)}{r^2},
\end{equation}
where $\theta$ is the polar angle, $\alpha_{{\rm He}^+}=9/2^5\approx 0.28$ a.u.\ is the polarizability of He$^+$ \cite{Kotani1951},  and $\eta$ is a  parameter 
describing the range of the polarization potential. 
 The polarization potential 
is constructed 
so 
 that $V_{\rm pol}(r\to\infty)=-\Efield \alpha_{{\rm He}^+} \cos\theta/r^2$ and $V_{\rm pol}(r\to0)=0$.
Finally, the full effective SAE potential including $V_{\rm pol}(\bm{r})$ reads
\begin{equation}\label{V2}
V_2(\bm{r})=V_{1}(\bm{r})+V_{\rm pol}(\bm{r}).
\end{equation}
The parameter $\eta$ is fixed 
so that two 
 times the polarizability of the single active electron matches 
 the polarizability  of He.

For He, we have also investigated a more complicated  SAE  potential of the form 
$V_3(r)=-[1+e^{-\zeta_1r}+\beta r(r-q)e^{-\zeta_2r}]/r$, 
where the parameters $\zeta_1$, $\zeta_2$, $\beta$, and $q$ are determined by the minimization of  the difference between the ground state single-electron density obtained by the SAE approximation and the ground state single-electron density obtained by the MCTDHF method. 
The difference between 
 the ionization rates obtained by using 
the potential $V_3(r)$ and the  ionization rates obtained by  the potential $V_1(r)$ defined in Eq.~\eqref{V1} was found to be small. 
Indeed, the relative difference 
is smaller than 4\% for fields $0.2\le \Efield\le 0.6$ a.u. We therefore conclude that the ionization rate is not sensitive to 
a small variation of the form of the SAE potential as long as the ionization potential is well reproduced.

The ionization rates in the SAE approximation are  obtained by calculating the 
lowest energy 
 eigenvalue $\epsilon'_{\rm SAE}$ of  the complex-rotated Hamiltonian 
$H'_{\rm SAE}(e^{i\Theta}\bm{r}')$ \cite{Reinhardt1982}, 
\begin{equation}\label{SAE_eigenvalueEq}
H'_{\rm SAE}(e^{i\Theta}\bm{r}')\psi'_{\rm SAE}(\bm{r}') = \epsilon'_{\rm SAE} \psi'_{\rm SAE}(\bm{r}'),
\end{equation} 
with $\Theta =0.2$. From the complex-valued eigenvalue $\epsilon'_{\rm SAE}$,
we obtain the scaled tunneling rate for a two-electron atom as
\begin{equation}\label{WSAE}
\WSAE' = -4\imag(\epsilon'_{\rm SAE}),
\end{equation}
and the 
unscaled original
  rate by the scaling relation \eqref{scalingrelation_rates}.
Note that there is an additional factor of 2 in Eq.~\eqref{WSAE}
originating from
 the indistinguishability of the two electrons.

In addition to the SAE approximation, 
we also
 discuss results obtained by the frozen core (FC) model of He \cite{Scrinzi2000},
\begin{align}\label{SAEfrozencoreHe1s2}
\Psi_{\rm FC} (\bm{r}_1,&\bm{r}_2,t)=
c_0(t)\Psi^{\rm He}_0(\bm{r}_1,\bm{r}_2)
\nonumber
\\
&+\frac{1}{\sqrt{2}}\left[\psi_0^{{\rm He}^+}(\bm{r}_1)\phi(\bm{r}_2,t)+\phi(\bm{r}_1,t)\psi_0^{{\rm He}^+}(\bm{r}_2)\right],
\end{align}
 where $\Psi^{\rm He}_0(\bm{r}_1,\bm{r}_2)$ is the ground state wave function of He, and $\psi_0^{{\rm He}^+}(\bm{r})$ is the ground state wave function of He$^+$. 
We can derive equations
 of motion for $c_0(t)$ and $\phi(\bm{r},t)$  by inserting $\Psi_{\rm FC} (\bm{r}_1,\bm{r}_2,t)$ into the TDSE \eqref{TDSE_unscaled}. 
  We obtain the ionization rate $\WFC$ by
 fitting the decreasing ground-state norm $|c_0(t)|^2$ to an exponential function $e^{-\WFC t-q}$.
 In the context of  
 static-field ionization of He, the ansatz \eqref{SAEfrozencoreHe1s2} was introduced by Scrinzi \cite{Scrinzi2000}, and similar models have been successfully used 
 in the numerical simulations of
   many-electron molecules in strong fields \cite{SpannerPatchkovskii2009,SPANNER201310,MajetyScrinzi2015,MajetyZielinskiScrinzi2015}.
 The wave function \eqref{SAEfrozencoreHe1s2} describes a single active electron with the ejected part of the wave function given by $\phi(\bm{r},t)$, 
 and includes the exchange interaction 
 of the ejected electron with the He$^+$ core,
 differently from the SAE model.
\section{Results}\label{Sec:Results}
\subsection{Comparison of MCTDHF rates and PPT rates}
\begin{figure}
\includegraphics[width=\figwidth]{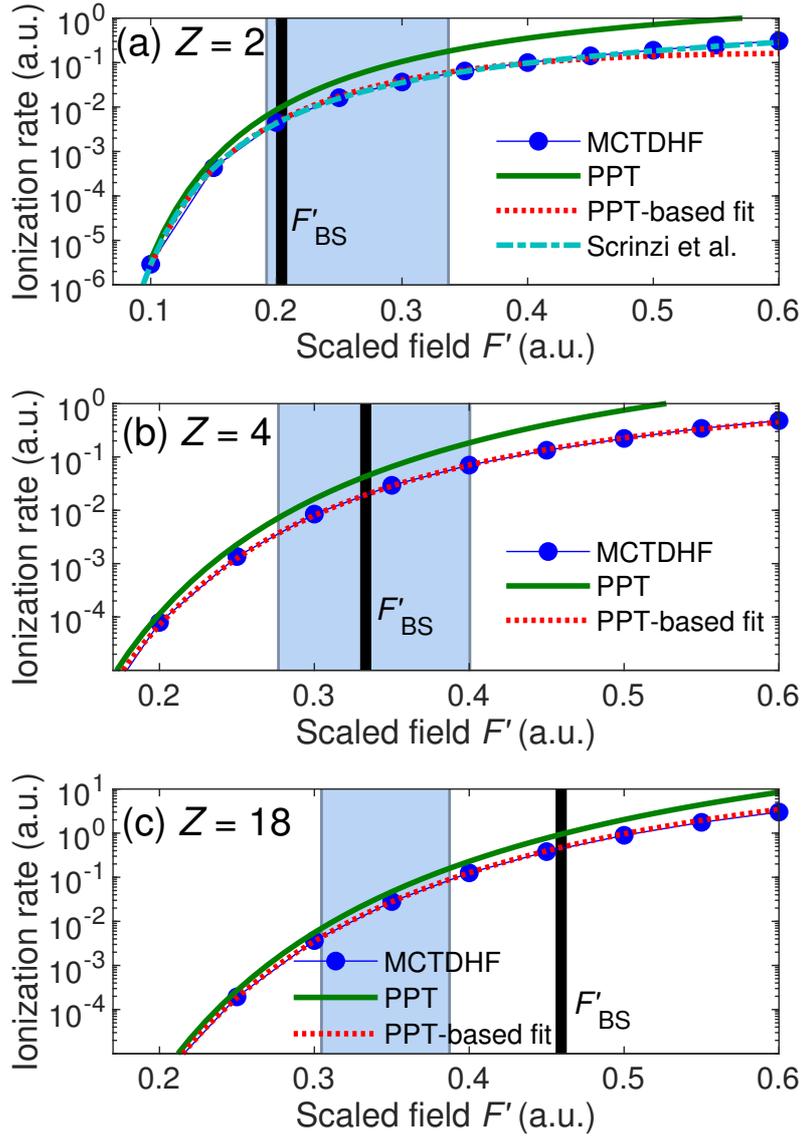}
\caption{\label{Fig1}Ionization rates $\W=Z^2\W'/4$ calculated with the MCTDHF method  for (a) He, (b) He-like Be ($Z=4$), and (c) He-like Ar ($Z=18$) are shown with filled circles. The ionization rates are shown as a function of the scaled field $F'=8F/Z^3$.  
Comparisons are made with the
 ionization rates calculated 
 using
  the PPT formula [solid line; see Eq.~\eqref{PPTrate}] 
  and
   the PPT-based fitting formula 
suggested in \cite{TongLin_2005} [dotted line; see Eq.~\eqref{modPPT}]. In panel (a), the dash-dotted line shows the results obtained by Scrinzi, Geissler and Brabec in 
\cite{ScrinziGeisslerBrabec1999}. The scaled barrier suppression field $F'_{\rm BS}$ is indicated with a 
vertical solid line. The blue shaded area indicates the interval where the ionization probability during one laser cycle 
$P_{\rm ion}$
[defined in Eq.~\eqref{Pioncycle}] 
is in the range between  10\% and 90\%.  
In terms of peak  laser field intensities, $F'=0.4$ a.u.\ corresponds to $5.6\times 10^{15}$ W/cm$^2$ for $Z=2$, 
$3.6\times 10^{17}$ W/cm$^2$ for $Z=4$, and $3.0\times 10^{21}$ W/cm$^2$ for $Z=18$.}
\end{figure}

In Fig.~\ref{Fig1}, we show the comparison 
between 
the 
ionization rates $\W$ calculated with the MCTDHF method and the ionization rates $\WPPT$ calculated with the PPT formula 
\cite{PPT_1966,V_Popruzhenko_2014,CiappinaPopruzhenko_2020}. The PPT formula 
employed in this study is given by   Eq.~\eqref{PPTrate} in Appendix ~\ref{PPTformula}. Note that the PPT rate
is defined with an additional  factor of 2,
originating from
 the fact that two electrons contribute to the total ionization rate.
For He 
($Z=2$),  a comparison is made with the rates $\W_{\rm SGB}$ calculated by Scrinzi, Geissler and Brabec \cite{ScrinziGeisslerBrabec1999} 
using the complex scaling method for a correlated two-electron wave function.

Also shown in Fig.~\ref{Fig1} are 
the results obtained by the 
PPT-based fitting  formula for the ionization rate 
  suggested by Tong and Lin \cite{TongLin_2005},
\begin{equation}\label{modPPT}
\WTong(\Efield) = \WPPT(\Efield) \Lambda_\alpha(F) ,
\end{equation}
where 
\begin{equation}\label{empirical_correctionPPT}
\Lambda_\alpha(\Efield)=\exp\left[-\alpha\frac{(Z-1)^2}{I_p}\frac{\Efield}{(2I_p)^{\frac{3}{2}}}\right]
\end{equation}
is an empirical correction factor, and $I_p$ is the ionization potential. For  the $I_p$ of He-like ions, we use the expression given in Eq.~\eqref{Ip_def} in Appendix \ref{PPTformula}.
We note that the expression \eqref{modPPT} should be viewed as a formula with which the ionization rates obtained 
by ab initio methods can be fitted.  
In \cite{TongLin_2005}, the parameter $\alpha$ in the  empirical correction factor 
$\Lambda_\alpha(F)$  was suggested to be 
$\alpha=7$ for He, but the values for other He-like ions were not given. We derive $\alpha$ by a 
least-squares fit of the 
PPT-based fitting formula $\WTong(\Efield)$ to the rate obtained by the MCTDHF method, using the ionization rates in the range
of  $0.2\le \Efield'\le 0.4$ a.u. 
We obtain $\alpha=7.12$ for $Z=2$, $\alpha=  7.14$ for $Z=4$, and $\alpha= 5.62$ for $Z=18$.
The 
applicability  of the PPT-based fitting formula \eqref{modPPT} has been confirmed
previously in the case of   strong-field ionization of H,
where
 the PPT-based fitting formula \eqref{modPPT} with $\alpha=6$ differs from accurate numerical
ionization rates  by less  than 10\% for 
the unscaled fields of 
 $\Efield<0.125$ a.u.\ \cite{TongLin_2005}.

Because $\Lambda_\alpha(\Efield)$ 
becomes exponentially small
as $\Efield$ increases,
the formula \eqref{modPPT} 
cannot be valid when $\Efield$ becomes too large,
 but, as can be seen in Fig.~\ref{Fig1}, it is remarkably accurate 
as long as the field 
strength  is in the tunneling regime as well as in the OBT regime below the field strength 
at 
$\Efield\approx2\EBS$, where the barrier-suppression field $\EBS$ is
defined as 
\begin{equation}\label{EBS_def}
\EBS = \frac{I_p^2}{4(Z-1)}.
\end{equation}
The quantity $Z-1$ 
represents 
 the residual charge 
with which  the ejected electron interacts. 
We regard that the field strength is in the tunneling regime when 
$\Efield<\EBS$ and that the field strength is in the OTB regime when $\Efield>\EBS$.
We show $\EBS$ in Fig.~\ref{Fig1} as a thick vertical line.
 The scaled barrier-suppression field $\Efield'_{\rm BS}=8\Efield_{\rm BS}/Z^3$ is defined according to the scaling relation, Eq.~\eqref{FieldScaling}.

In \cite{ZhangLanLu2014}, an improved 
empirical correction factor 
$\Lambda_{\alpha_1\alpha_2\alpha_3}(\Efield) = 
\exp(-\alpha_1\Efield^2/\EBS^2-\alpha_2\Efield/\EBS-\alpha_3)$ to the PPT formula with the three fitting  parameters $\alpha_1$, $\alpha_2$, $\alpha_3$  was proposed. 
However, we
 employ the original correction factor introduced in \cite{TongLin_2005},
 in which only one fitting parameter, $\alpha$, is used, for  its  simplicity.

In order to relate  the static-field rates 
to the probability of ionization in an oscillating laser field, we evaluate the ionization probability 
$P_{\rm ion}$ during one optical cycle of a laser field with 
the 
peak amplitude $\Efield$ and the 
angular frequency $\omega$ as
\begin{equation}\label{Pioncycle}
P_{\rm ion}(\Efield)=1-\exp\left[-\frac{1}{\omega}\int_0^{2\pi} \W(|\Efield \sin s|) \rmd s  \right],
\end{equation}
where $\omega=0.057$ a.u., corresponding to a wavelength of 800 nm, 
is chosen.
The expression \eqref{Pioncycle} for the ionization probability of  a He-like ion in a laser field  is valid under the assumption  of  a slowly varying field. 
In Fig.~\ref{Fig1}, the interval where $P_{\rm ion}$ fulfills $0.1\le P_{\rm ion}\le 0.9$ is indicated by 
the area shaded with 
light blue color.

We can see in Fig.~\ref{Fig1} that the standard PPT 
formula overestimates the ionization rates compared with the reference MCTDHF rates, even though the extent of the overestimation becomes smaller 
 for 
 the 
 larger values of $Z$. 
 On the other hand, the rates obtained by the  PPT-based fitting formula agree well with the MCTDHF rates. 
For $Z=2$, the relative 
deviation
 takes the values of 
  $|\WTong-\W|/\W<0.1$ 
  when   $\Efield'\le 0.4$ a.u.\ and $|\WTong-\W|/\W<0.5$ 
  when
    $\Efield'\ge 0.4$ a.u.
 For $Z=4$, we have  $|\WTong-\W|/\W<0.1$ 
 when
  $\Efield'\ge 0.25$ a.u. For $Z=18$, $|\WTong-\W|/\W<0.1$ when $0.25\le \Efield'\le 0.5$ a.u.\ and
 $|\WTong-\W|/\W<0.15$ when $\Efield'> 0.5$ a.u.
 This shows that the PPT-based fitting formula has a relative accuracy of 10\% 
 or better in the electric field  range where the ionization probability is substantially large  
 but below 
 the
 saturation 
 ($0.1<P_{\rm ion}<0.9$), corresponding to
  the range indicated by light blue color in Fig.~\ref{Fig1}.
 An accurate formula for the ionization rate in this 
 field strength
  range is crucial 
 for the simulation of the ion yields required in the intensity estimation scheme proposed in 
  \cite{Ciappinaetal2019,CiappinaPopruzhenko_2020}.

 In the case of  He, the MCTDHF rates and the rates $\W_{\rm SGB}$ calculated by Scrinzi, Geissler, and Brabec \cite{ScrinziGeisslerBrabec1999} are in good agreement with each other, 
 which confirms the validity of the MCTDHF method for calculating static-field ionization rates. The relative 
 deviation takes the values of
$|\W_{\rm SGB}-\W|/\W_{\rm SGB}<0.05$ when $\Efield'\le 0.5$ a.u.\ and $|\W_{\rm SGB}-\W|/\W_{\rm SGB}<0.08$ when $\Efield'>0.5$ a.u.
The small discrepancy between our results and the  results obtained in \cite{ScrinziGeisslerBrabec1999} can be ascribed to the difference in the method used 
in the calculations. 
In \cite{ScrinziGeisslerBrabec1999},  the ionization rates were obtained by the complex scaling 
of 
 a two-electron wave function expanded in terms of explicitly correlated basis functions. 
The rates for He obtained by the MCTDHF method are also in good agreement with the 
ionization rates reported by Parker et al.\ in \cite{ParkerArmstrongBocaTaylor2009}, where the TDSE was integrated in real time and the ionization rate was obtained by  monitoring the decay of the norm of the wave function in a sphere 
whose radius is 18 a.u.\ around the atomic core. For comparison, 
numerical values of the ionization rates for He obtained by the MCTDHF method are listed 
together with the ionization rates 
reported in  Ref.~\cite{ScrinziGeisslerBrabec1999} and 
Ref.~\cite{ParkerArmstrongBocaTaylor2009} in Appendix \ref{Sec:Herates_comparison}.

\begin{figure}
\includegraphics[width=\figwidth]{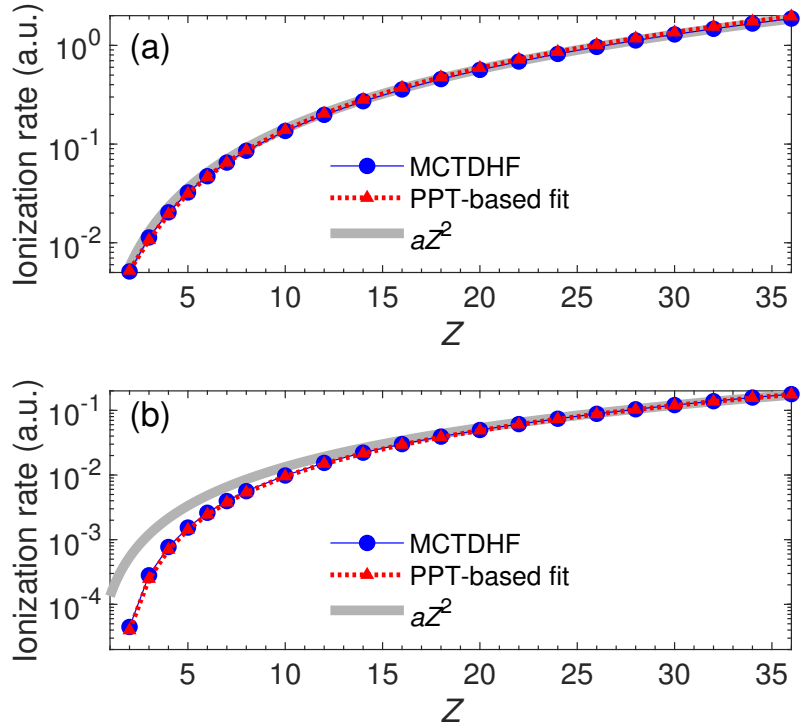}
\caption{\label{Fig2}(a) Ionization rate $\W=Z^2\W'/4$ at $\Efield=\EBS$ calculated with the MCTDHF method (blue solid circles), the rate $\WTong$ calculated with the 
PPT-based fitting formula (red triangles) and a fit of $\W$ to a quadratic function $\afit Z^2$ with $\afit=1.4\times 10^{-3}$ (thick solid gray line)  are shown as a function of $Z$. (b) Ionization rates evaluated at $\Efield=(2I_p)^{3/2}\times 0.05$, such that the reduced field $\Fred=0.05$ a.u. The coefficient $\afit$ in the quadratic fit is $\afit=1.3\times 10^{-4}$.
}
\end{figure}

To further assess the accuracy of the PPT-based fitting formula, we evaluate the ionization rate as a function of $Z$ at two 
different 
values of $\Efield$.
In Fig.~\ref{Fig2}(a), we show the ionization rate evaluated at the barrier-suppression field $\EBS$  for 
 $Z\le 36$. We compare the rate $\W(\Efield=\EBS)$
 calculated with the MCTDHF method and the rate $\WTong(\Efield=\EBS)$ calculated with the PPT-based fitting formula  defined
  by Eq.~\eqref{modPPT}.  For each value of $Z$, $\alpha$ is obtained by fitting 
  $\WTong(\Efield')$ to the ionization rates 
 $\W(\Efield')$ calculated by the MCTDHF method  for 
 the  scaled fields in the range
 of
  $0.2\le\Efield'\le0.4$ a.u. The numerical values of $\alpha$ are given in 
 Table~\ref{alphaTable}. 
 For  reference, we provide 
 numerical values of the ionization rates in the Supplemental material 
 \footnote{See Supplemental 
 material at [link inserted by PRA] for a text file containing the static-field ionization rates obtained by the MCTDHF method 
 for $2\le Z\le 36$}.
 
 The relative difference between $\W$ and $\WTong$,
 defined as $|\WTong(Z)-\W(Z)|/\W(Z)$,
  is less than 6\% for all $Z\ge2$. 
  It can also be seen in Fig.~\ref{Fig2}(a)
  that   the $Z$-dependence of the ionization rate $\W$ at $\Efield=\EBS$ 
  is quadratic  to a good approximation.
 This means that the 
 scaled rate $\W'(\EBS')$ is almost independent of $Z$ because
 $\W$ is scaled by $Z^2/4$ with respect to $\W'$ as shown in the relation \eqref{scalingrelation_rates}.
  A fit of $\W$ to the function $\W_{\rm fit}=\afit Z^2$
 gives $\afit \approx 1.4\times 10^{-3}$.
 As can be seen  in Fig.~\ref{Fig2}(a),  $\W_{\rm fit}=\afit Z^2$ agrees well with the rate obtained by the MCTDHF method. 
 For $Z$ in the range $2\le Z\le10$, 
 the relative error of the quadratic fit is 
 $|\W_{\rm fit}(Z)-\W(Z)|/\W(Z)<0.15$, and for $Z>10$,  it becomes $|\W_{\rm fit}(Z)-\W(Z)|/\W(Z)<0.05$.
 
 
 \begin{table}
\caption{\label{alphaTable}Parameter $\alpha$ 
adopted  in the PPT-based fitting formula \eqref{modPPT}.}
\begin{ruledtabular}
\begin{tabular}{m{\colw}m{\colw} | m{\colw}m{\colw} | m{\colw}m{\colw} | m{\colw}m{\colw}}
$Z$ & $\alpha$ & $Z$ & $\alpha$ & $Z$ & $\alpha$ & $Z$ & $\alpha$\\
\hline
    2&      7.12&        8  &     6.23 &        20&      5.57&       32&      5.41\\
    3&      7.52&        10&      6.00  &      22&      5.53&       34&       5.40\\
    4&      7.14&        12 &     5.88  &      24 &     5.49&       36&      5.39\\
    5&      6.84&        14 &     5.77  &      26 &     5.47 &      &         \\
    6&      6.58&        16  &    5.68  &      28 &     5.45 &       &         \\
    7&      6.38&        18  &    5.62 &       30  &    5.43 &        &         \\
\end{tabular}
\end{ruledtabular}
\end{table}

 In Fig.~\ref{Fig2}(b), we show the ionization rates evaluated at 
 $\Efield=(2I_p)^{3/2}\times0.05$ so
  that the reduced field $\Fred$, defined as
 \begin{equation}\label{defFred}
\Fred=\frac{\Efield}{(2 I_p)^{3/2}},
\end{equation}
equals $0.05$ a.u. The reduced field $\Fred$ appears in the PPT rate as 
$\exp[-2/(3\Fred)]$ [see Eq.~\eqref{PPTrate}], and therefore, 
 the PPT rate  varies rapidly with $\Fred$ in the tunneling regime. The particular value of
$\Fred=0.05$ a.u.\ was empirically proposed in  \cite{Ciappinaetal2019} as a threshold for  saturated ionization, beyond which atoms are completely ionized. The condition 
of 
$\Fred=0.05$ a.u.\ means that the 
field strength is  in the tunneling regime rather than in the over-barrier regime. 
For example, 
for $Z=2$, $\Efield=(2I_p)^{3/2}\times 0.05=0.12\;{\rm a.u.}<\EBS=0.20$ a.u., and, for $Z=10$, 
$\Efield=(2I_p)^{3/2}\times 0.05=41\;{\rm a.u.}<\EBS=54$ a.u.
As shown in Fig.~\ref{Fig2}(b), the PPT-based fitting formula works well also at the reduced  field  of $\Fred=0.05$ a.u. 
 Indeed, the deviations are sufficiently small in the entire range of $Z$, that is,
$|\WTong-\W|/\W<0.12$ for  $Z\le 12$ and $|\WTong-\W|/\W<0.03$ for  $Z> 12$.
However, it should be noted that the quadratic function fits well only 
in 
the large $Z$ above  
$\sim 20$, that is, (i) $|\W_{\rm fit}-\W|/\W>0.4$ for $Z<10$, 
(ii) $0.4>|\W_{\rm fit}-\W|/\W>0.1$ for $10\le Z<20$, and 
(iii) $|\W_{\rm fit}-\W|/\W<0.1$ for $Z\ge20$.

\subsection{Ionization rates by perturbation theory}
\begin{figure}
\includegraphics[width=\figwidth]{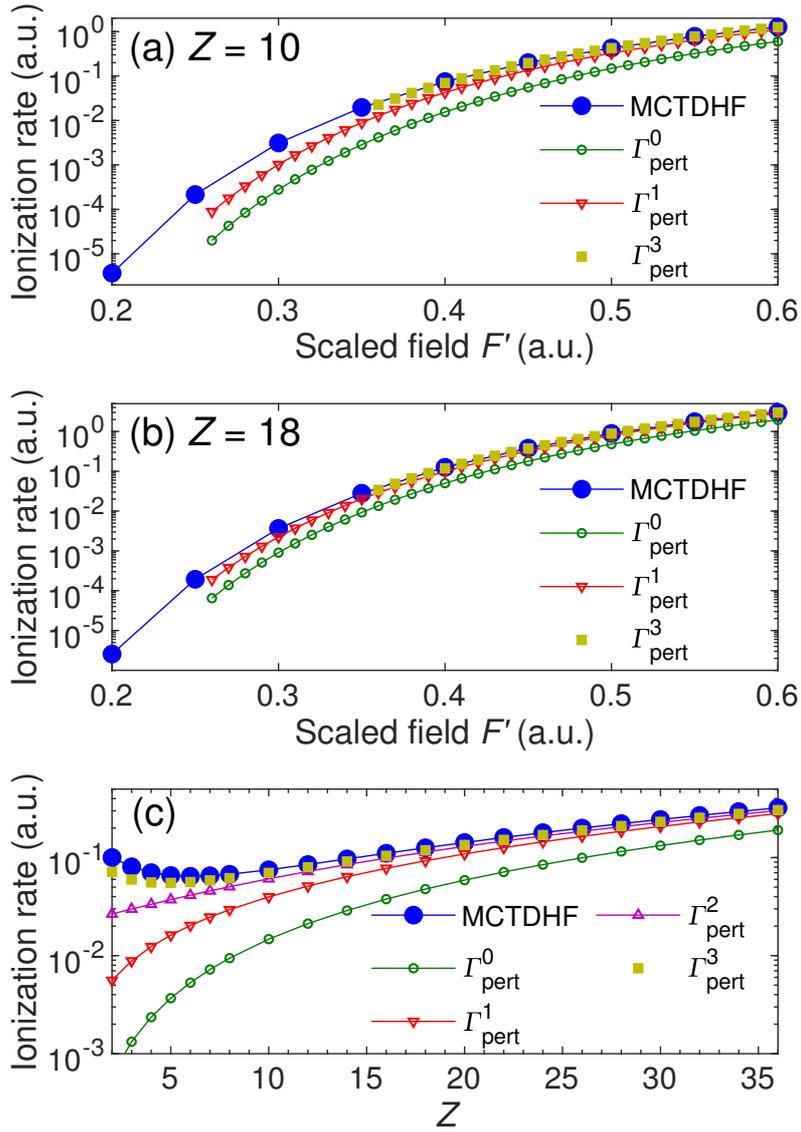}
\caption{\label{Fig3}Ionization rates  for (a) $Z=10$ and (b) $Z=18$ obtained by the MCTDHF method 
(filled circles) and those obtained by perturbation theory as
 a function of the 
scaled field $\Efield'=8\Efield/Z^3$. The perturbative rate 
$\Wpert^N$ contains corrections up to 
the 
order $1/Z^N$. In panel (c), we show the rates as a function of 
$Z$ at
 a fixed value $\Efield'=0.4$ a.u.\ of the scaled field.}
\end{figure}

In Fig.~\ref{Fig3}, we compare the ionization rates $\W$ calculated using the MCTDHF method and
 the ionization rates $\Wpert^N$ [see Eq.~\eqref{perttheory_W}] calculated by perturbation theory. The numerical values of the expansion coefficients $\lambda'_n$ 
 given by Eq.~\eqref{pertCoeffs}
  are shown in Appendix \ref{PertTheoryAppendix_subsec2}. 
Our numerical approach for solving the 
equations of perturbation theory, 
 described in Appendix \ref{PertTheoryAppendix_subsec2}, limits 
the range of the
 scaled field strengths to $\Efield'\ge 0.25$ a.u.\ for 
 the orders of $n=0$ and 1, and to
$\Efield'\ge 0.35$ a.u.\ for the orders of $2$ and 3.

We can see in Figs.~\ref{Fig3}(a) and (b) that the rate $\Wpert^{N=3}$ including  the third-order corrections agrees well with the rate calculated with the 
MCTDHF method. 
For both $Z=10$ and $Z=18$, the relative 
deviations are as small as  $|\Wpert^{N=3}-\W|/\W<0.07$ in the scaled field range of 
 $\Efield'\ge 0.4$ a.u. 
The values of $Z=10$ and $Z=18$ are selected to demonstrate the performance of 
perturbation theory for 
the 
large values of $Z\ge 10$, where 
the perturbation  to the third order
  is enough to
obtain ionization rates with a sufficiently small deviation from the reference MCTDHF rates.  
It can be seen in Figs.~\ref{Fig3}(a) and (b) that
  perturbation theory gives relatively accurate results for the ionization rate in both the tunneling and the OTB regimes (for $Z=10$, $\EBS'=0.43$ a.u., 
and for $Z=18$, $\EBS'=0.46$ a.u.). In the tunneling regime, where the ionization rate is exponentially sensitive to 
the variation in   the ionization potential, it is necessary to include   
sufficiently large orders in perturbation theory
so
 that an accurate value of the ground state energy is obtained. The results shown in Figs.~\ref{Fig3}(a) and (b) suggest that $N=3$ is large enough to obtain ionization rates 
with a relative deviation from the MCTDHF rates below 7\%  both in the tunneling and the OTB regimes.

In Fig.~\ref{Fig3}(c), we show the ionization rates as a function of $Z$, evaluated 
at the scaled field of  $\Efield'=0.4$ a.u. The deviation of the ionization rate calculated by the third order perturbation theory 
becomes
 smaller as $Z$ increases, that is, 
$|\Wpert^{N=3}-\W|/\W=0.29$ for $Z=2$, 
$|\Wpert^{N=3}-\W|/\W=0.09$ for $Z=8$, and
 $|\Wpert^{N=3}-\W|/\W<0.07$ for $Z\ge 10$. However,  the third-order rate 
 $\Wpert^{N=3}$ exhibits  a minimum in the rate 
at  $Z=5$ in a similar manner as  the MCTDHF rate exhibiting a minimum at $Z=6$, 
as can be seen in the curves plotted with filled circles and filled squares in Fig.~\ref{Fig3}(c). 
This minimum in the ionization rate appearing at a certain $Z$ value can be explained by  the asymptotic behavior of the 
PPT rate. At the fixed scaled field of $\Efield'=8\Efield/Z^3=0.4$ a.u., the PPT rate increases as $\WPPT\propto Z^2$ for large $Z$  ($Z>30$).  For the smaller values of $Z\le 7$, where the ionization rate is in the tunneling regime, $\WPPT$ decreases as $Z$  increases because  the  reduced field 
$\Fred\approx \frac{\Efield'}{8} \left(1-\frac{5}{4Z}\right)^{-\frac 3 2}$ decreases as 
$Z$ increases. Therefore, in the intermediate range, there must be a minimum in 
the ionization rate, which is expected to appear  approximately at the value of $Z$ where 
$\Efield' \approx \EBS'$. In the case of $ \Efield'=0.4$ a.u.,  $\Efield'<\EBS'$ when $Z\le 7$.

\subsection{Comparison of SAE approximation rates and MCTDHF rates}\label{subsec:SAE_MCTDHF_comp}
\begin{figure}
\includegraphics[width=\figwidth]{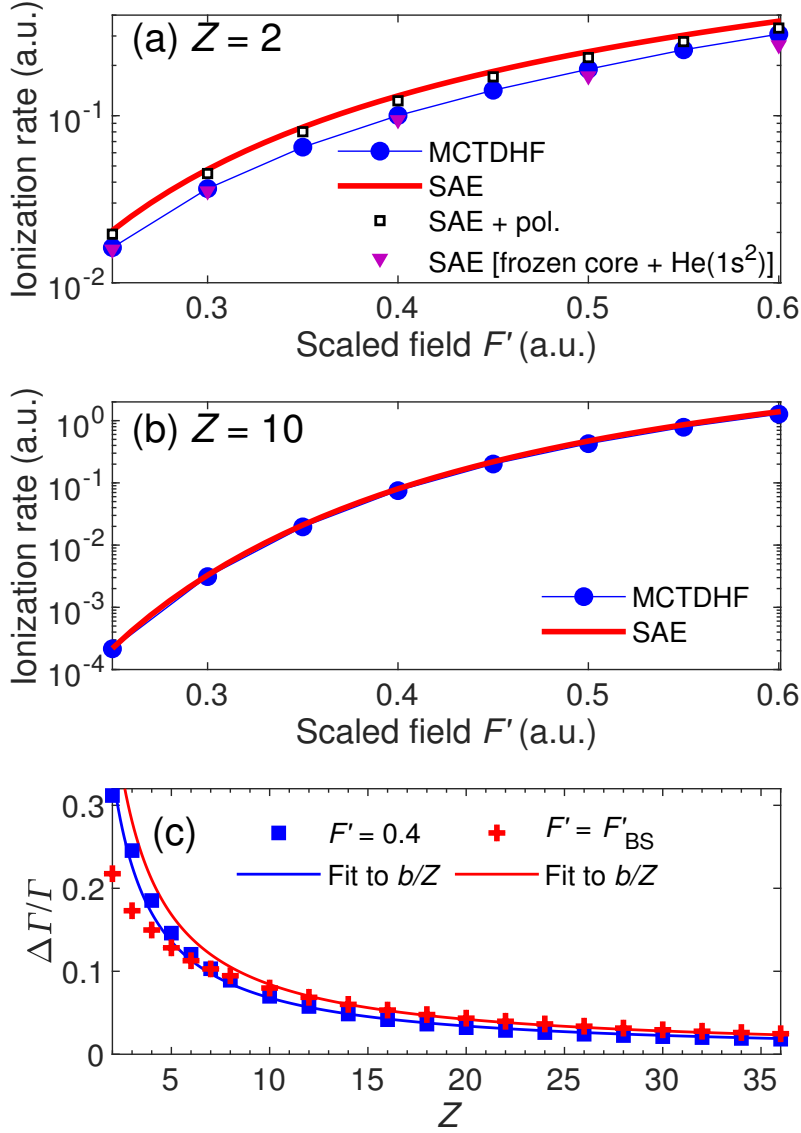}
\caption{\label{Fig4}Ionization rates $\W=Z^2\W'/4$ as a function of $\Efield'$ for (a) He and (b)  He-like Ne ($Z=10$), calculated with the MCTDHF method 
and  the
 SAE model, using potential $V_{1}$ [see Eq.~\eqref{V1}]. In (a) we additionally show the rates obtained with the SAE model including the polarization potential  
[see Eq.~\eqref{V2}], and the FC  model including a correlated He ground state wave function [see Eq.~\eqref{SAEfrozencoreHe1s2}]
In (c), we show the relative 
deviation $\Delta\W/\W =(\WSAE-\W)/\W$ of the rate $\WSAE$
 obtained with the SAE approximation  from the MCTDHF rate $\W$
  as a function of $Z$,
 evaluated at $\Efield'=0.4$ a.u.\ 
 and
  $\Efield=\EBS$. Also shown are fits to the function $b/Z$, with $b(\Efield'=0.4)=0.68$, and $b(\Efield=\EBS)=0.84$. In the case of 
 $\Efield=\EBS$, only $\Delta\W/\W$ values for $Z\ge 10$ were included in 
 the fit. }
\end{figure}

In Fig.~\ref{Fig4}, we show 
the comparison between the 
results obtained by the SAE approximation 
described in Sec.~\ref{subsec:SAE} and those obtained by the MCTDHF method. The numerical values of the  parameter $\zeta'$  are 
$\zeta'=2.1325$ for $Z=2$ and  $\zeta'=2.4489$ for $Z=10$.
For the polarization potential defined in Eq.~\eqref{Vpol}, we employ the value  $\eta=4.8713$ for He ($Z=2$), 
so that  twice the polarizability $\alpha_{\rm SAE}$ of 
the active electron equals the polarizability 
of He, $\alpha_{\rm He}=1.3832$ a.u.\ \cite{Yangetal2017}.  The numerical calculation of the SAE rates and the  calculation of the  SAE polarizability are described in 
Appendix \ref{AppendixSAE}. 

We can see in Fig.~\ref{Fig4}(a) that the SAE rate 
for He is larger than  the MCTDHF rate by around 30\% (31\% for $\Efield'= 0.4$ a.u.\ and 27\% for $\Efield'= 0.5$ a.u.). 
The agreement is not significantly improved when the  polarization potential is 
included in the  effective SAE potential, as defined in Eq.~\eqref{V2}. 
 On the other hand, the results obtained by the  FC model defined in  Eq.~\eqref{SAEfrozencoreHe1s2}
agree very well with the MCTDHF results 
($|\WFC-\W|/\W<0.06$ for $\Efield'\le 0.4$ a.u.). 
Because  the SAE model 
should give  the correct ionization potential 
in its  construction and the 
FC model,
in which the correlated ground state is included  in the wave function ansatz \eqref{SAEfrozencoreHe1s2}, should also give
 the correct ionization potential, the better agreement 
 of the ionization rate given by
  the FC model cannot be attributed to  the difference in 
   the ionization potential. 
   In addition, because the
 core polarization is not included in the FC model, the polarization of the He$^+$ 
 core 
 cannot be responsible 
  for the difference between the SAE results and the FC results. The difference between the FC model and the SAE model 
 can be ascribed to
  the exchange interaction between the ejected electron and the He$^+$ core 
  that is not included in the SAE model but in the FC model. 
  Indeed, the FC wave function \eqref{SAEfrozencoreHe1s2} is symmetric with respect to 
  the exchange of the two electron coordinates. Therefore, the disagreement between the SAE rate and the MCTDHF rate 
  is considered to originate from the exchange interaction. 
  The overestimation of the ionization rate given by the SAE approximation can be ascribed to the increase in the ionization probability originating from the absence of the exchange interaction. The importance of
   the exchange interaction 
   was reported before
    in the alignment dependence of the tunneling ionization rates in 
    CO$_2$, O$_2$, and N$_2$ \cite{MajetyScrinzi2015}.
In order to further prove that the exchange interaction can affect the ionization process, we have included in Appendix \ref{AppendixSAE_HF} a discussion 
on two different frozen-core Hartree-Fock models, that is, 
one in which the exchange symmetry is fulfilled and the other in which no exchange symmetry is considered.

Figure~\ref{Fig4}(b), in which the  MCTDHF rate and the SAE rate 
are compared for $Z=10$, shows that the relative difference is less than 10\% 
in the range of $0.25\le\Efield'\le 0.6$ a.u. 
In the case of $Z=18$ (not shown in Fig.~\ref{Fig4}), the relative 
difference is smaller than 6\% for $0.25\le\Efield'\le 0.6$ a.u., 
 suggesting that the 
effect of the exchange interaction decreases 
as  $Z$ increases. 

In order to quantify the effect of the exchange interaction, we define the 
relative deviation of the SAE rate from the MCTDHF rate
as 
\begin{equation}
\frac{\Delta\W}{\W}=\frac{\WSAE-\W}{\W},
\end{equation}
where $\W$ denotes the ionization rate calculated by the MCTDHF method,
and show $\Delta\W/\W$ as a function of $Z$ in Fig.~\ref{Fig4}(c). We evaluate the relative 
deviation at two different field strengths, $\Efield'=0.4$ a.u., and $\Efield=\EBS$. 
In both cases, the relative deviation $\Delta\W/\W$ 
decreases approximately as $1/Z$ as $Z$ increases, and the 
decrease  proportional to $1/Z$ is shown clearly 
 by the best fit curves having  the form of $b/Z$ with a constant factor of $b$.
 The deviation of the SAE rate from the MCTDHF rate caused by the exchange interaction is approximately 30\% in He, and 
 the deviation 
 decreases as $1/Z$ to about 2\% in He-like Kr.

\section{Summary and conclusions}
We have performed an  
investigation of the static-field  ionization rates of He-like ions for 
the 
nuclear charge numbers $Z$ up to $Z=36$, in both the tunneling and OTB regimes. We have 
made two major findings. The first is that both 
the PPT-based fitting formula by Tong and Lin \cite{TongLin_2005} [see Eq.~\eqref{modPPT} and Table~\ref{alphaTable}] and the third-order perturbation theory in $1/Z$ [see Eq.~\eqref{perttheory_W}] 
give small deviations from the rates calculated by the MCTDHF method. In the case of 
the PPT-based fitting formula,  
the 
relative deviation from the reference rates obtained by the MCTDHF method is
 about 10\%  
 when  $Z>2$ and  $\Efield\le0.075 Z^3$ a.u., 
 corresponding to scaled fields $\Efield'\le 0.6$ a.u. 
  In the case of the ionization rates calculated with third-order perturbation theory, 
 the relative deviation from the MCTDHF rates is 7\% 
  for  $Z\ge 10$.
 In practice, the PPT-based fitting formula with the 
 parameter $\alpha$ whose values are  given  in Table~\ref{alphaTable} 
 provides reasonably good estimates of the ionization rate and will help
  researchers 
 estimate the ion yields in
  the intensity measurement scheme proposed in \cite{Ciappinaetal2019,CiappinaPopruzhenko_2020}.
 
 The second finding is that the exchange interaction in 
 the 
 tunneling and OTB barrier ionization reduces the ionization rate by an amount proportional to $1/Z$, ranging from a reduction of  30\% for He 
 ($Z=2$) to about 2\% for He-like Kr 
 ($Z=36$). 
 This demonstrates the importance of the exchange interaction 
 in the 
 strong-field ionization of He-like ions, and shows how the relative strength of the exchange 
 interaction decreases as the nuclear charge number 
 $Z$ increases.

\begin{acknowledgments}
This research was supported by  JSPS KAKENHI grants no.\ JP18K05024 and  
no.\ JP15H05696, 
a JSPS Invitational Fellowship for Research in Japan (no.\ S18108), 
 and a travel grant  from the Czech Academy of Sciences (no.\ VAJVA-19-04).
M. F. C. acknowledges the project Advanced research using high intensity laser produced photons and particles (CZ.02.1.01/0.0/0.0/ 16\_019/0000789) from European Regional Development Fund (ADONIS), the Spanish Ministry MINECO (National Plan 15 Grant: FISICATEAMO No.\ FIS2016-79508-P, SEVERO OCHOA No.\ SEV-2015-0522, FPI), European Social Fund, Fundaci\'o Cellex, Generalitat de Catalunya (AGAUR Grant No.\ 2017 SGR 1341 and CERCA/ Program), ERC AdG NOQIA, and 
the National Science Centre, Poland-Symfonia Grant No.\ 2016/20/W/ST4/ 00314. 
\end{acknowledgments}

\appendix

\section{Numerical implementation of the MCTDHF method}\label{App_MCTDHF}
The numerical method used for the solution of the MCTDHF equations of motion  is the same as that presented in \cite{LotstedtKatoYamanouchi2017,LotstedtSzidarovszkyetal2020}. 
We 
describe 
 the essential points below.
The MCTDHF rates are obtained with $M=10$  spatial  orbitals for He. The orbitals  can initially (at $t'=0$) be characterized as $4s3p3p'$. For $Z>2$, we employ $M=6$ ($3s3p$).  
The spatial orbitals $\phi'_j(\bm{r}',t')$ are written as 
\begin{equation}\label{MCTDHForbitalexpansion}
\phi'_j(\bm{r}',\omega,t')=\sum_{\ell=0}^{\ell_{\rm max}} \frac{f_{\ell j}(r',t')}{r'}Y_{\ell, m_j}(\theta',\varphi'),
\end{equation}
where $Y_{\ell, m}(\theta',\varphi')$ are spherical harmonics and  $\ell_{\rm max}=10$. 
The magnetic quantum number $m_j$ is fixed for each orbital. The radial functions $f_{\ell j}(r')$ are discretized  on a non-uniform mesh  with $r'< 60$ a.u.\ and  
an 
asymptotic mesh width $\Delta r' =0.2$ a.u.\ at large $r'>5$ a.u. The grid points at smaller $r'$ are distributed such that the error in the ground state energy $\varepsilon_0({\rm He}^+)$ 
of He$^+$ is minimized. We obtain $\varepsilon_0({\rm He}^+)=-2+2\times 10^{-9}$ a.u. For the ground state of He (two electrons), we obtain with $M=10$ spatial orbitals  
a value of $\varepsilon_0({\rm He})=-2.8990$ a.u., which is $0.2\%$ larger than the exact, non-relativistic ground state energy 
$\varepsilon_0^{\rm exact}({\rm He})\approx-2.9037$ a.u.\ \cite{NakashimaNakatsuji2007}.

The time evolution is calculated by a predictor-corrector method similar to that presented in \cite{Haxton2011}, with a time step of $\delta t' =5\times 10^{-3}$ a.u.  Further details can be found in 
Appendix E of \cite{LotstedtKatoYamanouchi2017}.

\section{Perturbation theory}
\subsection{Equations for $\chi'_n$ and $\epsilon'_n$}\label{PertTheoryAppendix_subsec1}
 The 
equations for perturbation theory up to order three in $1/Z$ are given as
 \begin{equation}\label{def_chi0_perttheory}
 \chi'_0(\bm{r}'_1,\bm{r}'_2)=\xi'_0(\bm{r}'_1)\xi'_0(\bm{r}'_2),
 \end{equation} 
 \begin{equation}\label{xi0_perttheory}
 h'_0\xi'_0(\bm{r}')=\frac{\epsilon'_0}{2} \xi'_0(\bm{r}'),
 \end{equation}
 where $h'_0$ is the complex-rotated single-electron Hamiltonian for H-like He,
 \begin{equation}
 h'_0=-\frac{e^{-2 i\Theta}}{2}\frac{\partial^2}{\partial \bm{r}'^2} -\frac{2e^{-i\Theta}}{r'} + e^{i\Theta}F'z',
 \end{equation}
 \begin{equation}
 \epsilon'_1 = \Big(\chi'_0\Big|\frac{2e^{-i\Theta}}{r'_{12}}\Big|\chi'_0\Big),
 \end{equation}
 \begin{align}\label{xi1_eq}
 &\Big[\sum_{i=1,2}h'_0(e^{i\Theta}\bm{r}'_i)-\epsilon'_0\Big]\chi'_1(\bm{r}'_1,\bm{r}'_2)
 \nonumber
 \\
 &=
 \left(\epsilon'_1-\frac{2e^{-i\Theta}}{r'_{12}}\right) 
 \chi'_0(\bm{r}'_1,\bm{r}'_2),
 \end{align}
 \begin{equation}
 \epsilon'_2=\Big(\chi'_0\Big|\frac{2e^{-i\Theta}}{r'_{12}}\Big|\chi'_1\Big),
 \end{equation}
  \begin{align}\label{xi2_eq}
 &\Big[\sum_{i=1,2}h'_0(e^{i\Theta}\bm{r}'_i)-\epsilon'_0\Big]\chi'_2(\bm{r}'_1,\bm{r}'_2)
 \nonumber
 \\
 &=
 \left(\epsilon'_1-\frac{2e^{-i\Theta}}{r'_{12}}\right) 
 \chi'_1(\bm{r}'_1,\bm{r}'_2) +\epsilon'_2\chi'_0(\bm{r}'_1,\bm{r}'_2),
 \end{align}
 and finally
\begin{equation}\label{epsilon3_perttheory}
 \epsilon'_3=\Big(\chi'_0\Big|\frac{2e^{-i\Theta}}{r'_{12}}-\epsilon'_1\Big|\chi'_2\Big).
 \end{equation}
 
 The inner product $(\chi|\tilde{\chi})$ between two complex-rotated wave functions $\chi$ and $\tilde{\chi}$ is denoted by 
 round 
 brackets,
  and is defined 
 as
 \begin{equation}
(\chi|\tilde{\chi})= \int \rmd^3 r'_1 \rmd^3 r'_2\chi(\bm{r}'_1,\bm{r}'_2)\tilde{\chi}(\bm{r}'_1,\bm{r}'_2),
\end{equation}
without the complex conjugate on the left-hand wave function $\chi$ \cite{MoiseyevCertainWeinhold1978,BengtssonLindrothSelsto2008}.

To ensure that the wave function is consistently normalized to 1 at each order of $1/Z$, we should normalize $\chi'_0$ as $(\chi'_0|\chi'_0)=1$. We also have to project out the $\chi'_0$-component in $\chi'_1$ such that
$(\chi'_0|\chi'_1)=0$ is satisfied after solving Eq.~\eqref{xi1_eq}. Similarly, after solving Eq.~\eqref{xi2_eq}, we have to adjust $\chi'_2$ according to 
$\chi'_2\to\chi'_2-\kappa\chi'_0$ with $\kappa=(\chi'_0|\chi'_2)+(\chi'_1|\chi'_1)/2$, such that $(\chi'_0|\chi'_2)=-(\chi'_1|\chi'_1)/2$ is satisfied.

\subsection{Numerical values of $\lambda'_n$}\label{PertTheoryAppendix_subsec2}
\begin{figure}
\includegraphics[width=\figwidth]{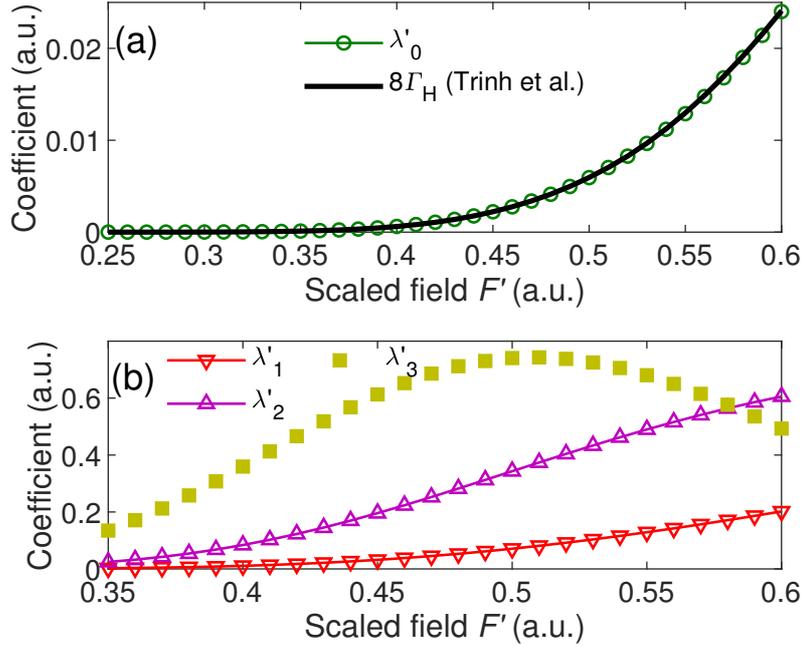}
\caption{\label{FigAppendix1} (a) Zeroth-order expansion coefficent $\lambda'_0$ and eight times the ionization rate $\WH$ of hydrogen, taken from 
Trinh et al.\ \cite{TrinhTolstikhinMadsenMorishita2013}. (b) Expansion coefficients $\lambda'_n$ for 
the 
orders $n=1$, 2, and 3.}
\end{figure}
 Equation~\eqref{xi0_perttheory} is a single-particle equation, and is solved by expanding $\xi'_0(\bm{r}')$ in the same way as shown in Eq.~\eqref{MCTDHForbitalexpansion}, 
 using a radial grid with $\Delta r=3\times 10^{-4}$ a.u., $\ell_{\rm max}=5$, and $m=0$. This approach is limited to scaled fields $\Efield'$ larger than $0.25$ a.u. To obtain  $\xi'_0$ and 
 $\epsilon'_{0,1}$ at smaller values of $\Efield'$ requires more sophisticated methods, such as 
 those in which multiple-precision arithmetic is employed 
 (see Appendix B of Ref.~\cite{TrinhTolstikhinMadsenMorishita2013}).
We solve the Eqs.~\eqref{xi1_eq} and \eqref{xi2_eq} by expanding the $n$th-order wave function $\chi'_n$ as
\begin{align}
\chi'_n(\bm{r}'_1,\bm{r}'_2)=\sum_{\ell,\ell'=0}^{\ell_{\rm max}} \sum_{m=-m_{\rm max}}^{m_{\rm max}} 
&\frac{C_{\ell\ell'm}(r'_1,r'_2)}{r'_1 r'_2}
\nonumber
\\
\times & Y_{\ell,-m}(\theta'_1,\varphi'_1)Y_{\ell',m}(\theta'_2,\varphi'_2),
\end{align}
and employ a grid  with $\Delta r' =0.2$ a.u., $r'<30$ a.u., $m_{\rm max}=2$, and $\ell_{max}=5$. We define $Y_{\ell,m}=0$ when $|m|>\ell$.
The electron-electron interaction term $1/r'_{12}$ is dealt with by the standard expansion in spherical harmonics \cite{Jackson1998}.
Note that both $\chi'_1$ and  $\chi'_2$ are correlated two-electron wave functions, and Eqs.~\eqref{xi1_eq} and \eqref{xi2_eq} are therefore considerably more difficult to 
be solved 
 than 
Eq.~\eqref{xi0_perttheory}. For $\epsilon_{2,3}$, our approach 
is limited to scaled fields $\Efield'\ge 0.35$ a.u.

In Fig.~\ref{FigAppendix1}, we show the expansion coefficients $\lambda'_n$ ($n=0,1,\ldots,3$)  which are used in Eq.~\eqref{perttheory_W} for the evaluation of the ionization rate by perturbation theory. 
According to Eq.~\eqref{ExactScaling_WH}, we have $\lambda'_0(\Efield')=8\WH(\Efield'/8)$, where $\WH$ is the ionization rate for H, and we therefore show also $\WH$ 
(taken from \cite{TrinhTolstikhinMadsenMorishita2013}) in Fig.~\ref{FigAppendix1}(a) for comparison. The agreement between our results for $\lambda'_0$ and the ionization rates for H presented in \cite{NicolaidesThemelis1992,Karlsson_1992,TrinhTolstikhinMadsenMorishita2013} is better than 1\% for $\Efield\ge 0.3$ a.u.

\section{PPT formula}\label{PPTformula}
The PPT formula employed in this paper reads \cite{PPT_1966,V_Popruzhenko_2014,CiappinaPopruzhenko_2020}
\begin{equation}\label{PPTrate}
\WPPT= 2C^2_\nu I_p \widetilde{F}^{1-2 \nu} \exp\left(-\frac{2}{3 \Fred}\right),
\end{equation}
where
\begin{equation}
\nu=\frac{Z-1}{\sqrt{2 I_p}} 
\end{equation}
is the effective principal quantum number defined in terms of the ionization potential $I_p$,   
the reduced field $\Fred$ is defined in Eq.~\eqref{defFred},
and
\begin{equation}
C^2_\nu=\frac{2^{4\nu-1}}{\nu\Gamma(\nu+1)\Gamma(\nu)}.
\end{equation}
Note the additional factor of 2  in the expression \eqref{PPTrate} which accounts for the contribution of two electrons.

In order to have a consistent definition of the ionization potential $I_p$ for all values of $Z$, we employ the expansion in inverse powers of $Z$ of the ground state energy 
$\varepsilon_0(Z)$
of a He-like ion given in \cite{SandersScherr1969},
\begin{equation}\label{invZexpansion}
\varepsilon_0(Z)=Z^2\sum_{n=0}^{25} \frac{a_n}{ Z^{n}}.
\end{equation} 
The coefficients $a_n$ in Eq.~\eqref{invZexpansion} are taken from table XII in \cite{SandersScherr1969}. 
The first two coefficients $a_0= -1$ and $a_1=5/8$ are known analytically, but $a_n$ for $n>1$ have to be calculated numerically.
The ionization potential is calculated as
\begin{equation}\label{Ip_def}
I_p(Z)=-\frac{Z^2}{2}-\varepsilon_0(Z).
\end{equation}
Compared to the exact nonrelativistic ground state energy of He, $\varepsilon_0^{\rm exact}(Z=2)\approx -2.9037$ a.u.\ \cite{NakashimaNakatsuji2007}, 
the expression \eqref{invZexpansion} gives $\varepsilon_0(2)-\varepsilon_0^{\rm exact}(2)\approx 10^{-7}$ a.u., while for $Z=10$ we have 
$\varepsilon_0^{\rm exact}(Z=10)\approx -93.9068$ a.u.\ \cite{NakashimaNakatsuji2007}  and 
$\varepsilon_0(10)-\varepsilon_0^{\rm exact}(10)\approx 5\times 10^{-8}$. This shows that the formula \eqref{invZexpansion} has a 
relative accuracy of $[\varepsilon_0(Z)-\varepsilon_0^{\rm exact}(Z)]/|\varepsilon_0^{\rm exact}(Z)|<4\times 10^{-8}$.
The experimental ionization potentials are $I_p^{\rm exp}(Z=2) \approx 0.9036$ a.u.\ and $I_p^{\rm exp}(Z=10) \approx 43.9451$ a.u.\ \cite{NIST_ASD2019}; the differences 
$I_p^{\rm exp}(2)-I_p(2)\approx -1.5\times 10^{-4}$ a.u.\ and  $I_p^{\rm exp}(10)-I_p(10)\approx 0.038$ a.u.\ can be attributed to corrections to the energy levels 
originating from the finite nuclear mass and relativistic effects.

\section{Numerical values of ionization rates for He}
\label{Sec:Herates_comparison}
In this Appendix, we list in Table \ref{Herates_numericalvalues} the values of the ionization rates $\W$ obtained in the present work by the MCTDHF method, together with the ionization rate $\W_{\rm SGB}$ calculated by Scrinzi, Geissler, and Brabec \cite{ScrinziGeisslerBrabec1999} and the ionization rate $\W_{\rm PABT}$ calculated by 
Parker, Armstrong, Boca, and Taylor \cite{ParkerArmstrongBocaTaylor2009}.
The relative deviations from the MCTDHF rates are $|\W_{\rm SGB}-\W|/\W_{\rm SGB}<0.08$ 
and $|\W_{\rm PABT}-\W|/\W_{\rm PABT}<0.04$ 
in the range of $\Efield$ shown in Table \ref{Herates_numericalvalues}.

\begin{table}
\caption{\label{Herates_numericalvalues}
Static-field ionization rates $\W$ of He obtained by different approaches.}
\begin{ruledtabular}
\begin{tabular}{c|c|c|c}
$\Efield/{\rm a.u.}$& $\W/{\rm a.u.}$ (MCTDHF)& 
$\W_{\rm SGB}/{\rm a.u.}$ \cite{ScrinziGeisslerBrabec1999}& 
$\W_{\rm PABT}/{\rm a.u.}$ \cite{ParkerArmstrongBocaTaylor2009}\\
\hline
    0.10&     $2.91\times 10^{-6}$	&         $2.88\times 10^{-6}$	&       	$2.9391\times 10^{-6}$ 	\\
    0.15&     $4.32\times 10^{-4}$	&         $4.23\times 10^{-4}$	&      	$4.2913\times 10^{-4}$ 	\\
    0.20&     $4.49\times 10^{-3}$	&         $4.31\times 10^{-3}$	&     	$4.3347\times 10^{-3}$ 	\\
    0.25&     $1.62\times 10^{-2}$	&         $1.57\times 10^{-2}$	&     	$1.5793\times 10^{-2}$ 	\\
    0.30&     $3.66\times 10^{-2}$	&         $3.56\times 10^{-2}$	&    		$3.5857\times 10^{-2}$   	\\
    0.35&     $6.48\times 10^{-2}$	&         $6.33\times 10^{-2}$	&       							\\
    0.40&     $1.00\times 10^{-2}$	&         $9.77\times 10^{-2}$	&       							\\
    0.45&     $1.42\times 10^{-1}$	&         $1.38\times 10^{-1}$	&       							\\
    0.50&     $1.90\times 10^{-1}$	&         $1.83\times 10^{-1}$	&       							\\
    0.55&     $2.48\times 10^{-1}$	&         $2.33\times 10^{-1}$	&       							\\
    0.60&     $3.08\times 10^{-1}$	&         $2.87\times 10^{-1}$	&       							\\
\end{tabular}
\end{ruledtabular}
\end{table}

   
%

\section{Numerical implementation of the single-active electron approximation}\label{AppendixSAE}
We solve the
 eigenvalue equation \eqref{SAE_eigenvalueEq} for the SAE rate 
 by expanding the wave function in the same manner as shown in Eq.~\eqref{MCTDHForbitalexpansion}, 
but using a smaller $\Delta r' = 3\times 10^{-4}$ a.u.\ for the radial grid, $m=0$, and $\ell_{\rm max}=5$.

In the frozen core model defined in Eq.~\eqref{SAEfrozencoreHe1s2}, we adopt
 the expansion \eqref{MCTDHForbitalexpansion} for $\phi(\bm{r},t)$, and 
we employ
 a MCTDHF wave function ($M=10$)  as defined in Eq.~\eqref{MCTDHF_def} for $\Psi^{\rm He}_0(\bm{r}_1,\bm{r}_2)$.

Within the SAE approximation, the polarizability $\alpha_{\rm SAE}$ is calculated as
\begin{equation}
\alpha_{\rm SAE}=-2\langle\psi^0_{\rm SAE}|z|\psi^1_{\rm SAE}\rangle,
\end{equation}
where $\psi^0_{\rm SAE}$ is the ground state  of the field-free SAE Hamiltonian $H_{\rm SAE}(\Efield=0)$, 
\begin{equation}
H_{\rm SAE}(\Efield=0)\psi^0_{\rm SAE}=\varepsilon_0\psi^0_{\rm SAE},
\end{equation}
and $\psi^1_{\rm SAE}$ solves 
\begin{equation}
[H_{\rm SAE}(\Efield=0)-\varepsilon_0]\psi^1_{\rm SAE}=-\left[z+\frac{V_{\rm pol}(\bm{r})}{\Efield}\right]\psi^0_{\rm SAE}.
\end{equation}

The parameter $\eta$ in Eq.~\eqref{Vpol} is adjusted so that
\begin{equation}
 2\alpha_{\rm SAE}=\alpha_{\rm He},
 \end{equation}
 where $\alpha_{\rm He}=1.3832$ a.u.\  is the polarizability of He \cite{Yangetal2017}.
 
\section{Frozen-core Hartree-Fock model with and without symmetrization}\label{AppendixSAE_HF}
\begin{figure}
\includegraphics[width=\figwidth]{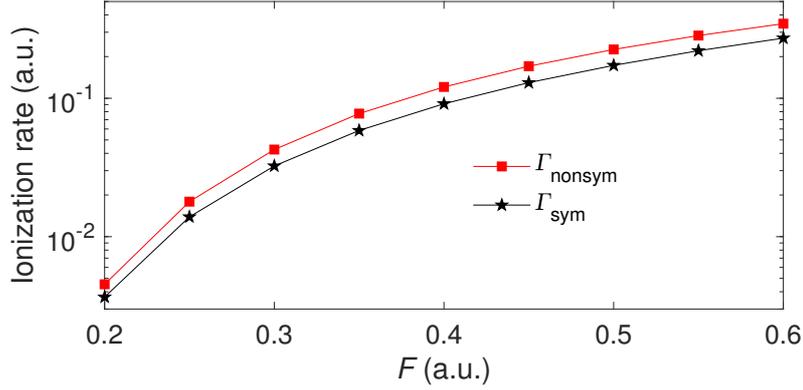}
\caption{\label{FigAppendix2}Static-field ionization rate $\W_{\rm sym}$  of He calculated by the Hartree-Fock model 
defined in Eq.~\eqref{SAE_HF_sym}
where exchange symmetry is fulfilled  and the ionization rate $\W_{\rm nonsym}$  calculated by the Hartree-Fock model 
defined in Eq.~\eqref{SAE_HF_nonsym}
where no exchange symmetry is considered.}
\end{figure}
In order to  corroborate  that the exchange interaction can influence the ionization rate, we consider two 
different 
wave function models of Hartree-Fock type for He. In the first, the two-electron wave function is written as 
\begin{equation}\label{SAE_HF_sym}
\Psi_{\rm HF(sym)}(\bm{r}_1,\bm{r}_2,t)=\frac{1}{\sqrt{2}}
\left[\psi^{{\rm HF}}_{1s}(\bm{r}_1)\phi(\bm{r}_2,t)+\phi(\bm{r}_1,t)\psi^{{\rm HF}}_{1s}(\bm{r}_2)\right],
\end{equation}
where $\psi^{{\rm HF}}_{1s}(\bm{r})$ is the Hartree-Fock ground state 
of the 
 1$s$ orbital of He, and $\phi(\bm{r},t)$ is 
a time-dependent wave function describing the ejected electron with initial condition 
$\phi(\bm{r},t=0)=\psi^{{\rm HF}}_{1s}(\bm{r})/\sqrt{2}$.
The wave function defined in Eq.~\eqref{SAE_HF_sym} has the correct
symmetry with respect to 
the
exchange of the electron 
coordinates.
 
In the second model, we take a wave function of the form
\begin{equation}\label{SAE_HF_nonsym}
\Psi_{\rm HF(nonsym)}(\bm{r}_1,\bm{r}_2,t)=\psi^{{\rm HF}}_{1s}(\bm{r}_1)\phi(\bm{r}_2,t),
\end{equation}
which is not symmetric with respect to
the exchange of the two electron coordinates. The initial condition for $\phi(\bm{r},t)$ 
in Eq.~\eqref{SAE_HF_nonsym} is $\phi(\bm{r},t=0)=\psi^{{\rm HF}}_{1s}(\bm{r})$.

Both models \eqref{SAE_HF_sym} and \eqref{SAE_HF_nonsym} have the same ground state, that is, the Hartree-Fock ground state, 
$\Psi_{\rm HF(sym)}^0(\bm{r}_1,\bm{r}_2)=\Psi_{\rm HF(nonsym)}^0(\bm{r}_1,\bm{r}_2)=\psi^{{\rm HF}}_{1s}(\bm{r}_1)\psi^{{\rm HF}}_{1s}(\bm{r}_2)$, and therefore, the same ionization potential. The only difference is the  absence of exchange symmetry in 
Eq.~\eqref{SAE_HF_nonsym}. Because the ionization potential of the models \eqref{SAE_HF_sym} and \eqref{SAE_HF_nonsym}
is too large, $I_p^{\rm HF}=0.9169$ a.u.\ compared to the accurate value 
of $I_p=0.9037$ a.u.\ given by Eq.~\eqref{Ip_def}, 
the resulting ionization rates are not quantitatively accurate, but the two models can be used to demonstrate the effect of the exchange interaction during 
the ionization.

The ionization rates for He obtained by the models \eqref{SAE_HF_sym} and \eqref{SAE_HF_nonsym} are shown in 
Fig.~\ref{FigAppendix2}. The numerical calculation was performed in the same way as described in Appendix~\ref{AppendixSAE}. In 
the non-symmetric model Eq.~\eqref{SAE_HF_nonsym}, the  ionization rate is multiplied with a factor of 2 to account for
the indistinguishability of the electrons.

We can see in Fig.~\ref{FigAppendix2} that the ionization rate $\W_{\rm nonsym}$ for the non-symmetric model is larger than the 
ionization rate $\W_{\rm sym}$ for the symmetric model. 
The relative difference given by $(\W_{\rm nonsym}-\W_{\rm sym})/\W_{\rm sym}$
is $0.32$ at $\Efield=0.4$ a.u.\
and that at $\Efield=0.5$ a.u.\ is $0.30$. These values of the relative difference are similar 
to the values of the relative difference 
$(\W_{\rm SAE}-\W)/\W$
between the rate $\W_{\rm SAE}$  obtained by the SAE approximation and the rate $\W$
obtained by the MCTDHF method,  where we obtain
$(\W_{\rm SAE}-\W)/\W=0.31$ at $\Efield= 0.4$ a.u.\ and 
$(\W_{\rm SAE}-\W)/\W=0.27$ at $\Efield= 0.5$ a.u.,
as shown in 
Fig.~\ref{Fig4}(a) in Sec.~\ref{subsec:SAE_MCTDHF_comp}.
 Because the only difference between the two wave functions \eqref{SAE_HF_sym} and \eqref{SAE_HF_nonsym} is the exchange symmetry, we conclude that inclusion of the exchange symmetry reduces the ionization 
rate by approximately 30\% also in the case of Hartree-Fock type wave functions.


\begin{thebibliography}{70}%
\makeatletter
\providecommand \@ifxundefined [1]{%
 \@ifx{#1\undefined}
}%
\providecommand \@ifnum [1]{%
 \ifnum #1\expandafter \@firstoftwo
 \else \expandafter \@secondoftwo
 \fi
}%
\providecommand \@ifx [1]{%
 \ifx #1\expandafter \@firstoftwo
 \else \expandafter \@secondoftwo
 \fi
}%
\providecommand \natexlab [1]{#1}%
\providecommand \enquote  [1]{``#1''}%
\providecommand \bibnamefont  [1]{#1}%
\providecommand \bibfnamefont [1]{#1}%
\providecommand \citenamefont [1]{#1}%
\providecommand \href@noop [0]{\@secondoftwo}%
\providecommand \href [0]{\begingroup \@sanitize@url \@href}%
\providecommand \@href[1]{\@@startlink{#1}\@@href}%
\providecommand \@@href[1]{\endgroup#1\@@endlink}%
\providecommand \@sanitize@url [0]{\catcode `\\12\catcode `\$12\catcode
  `\&12\catcode `\#12\catcode `\^12\catcode `\_12\catcode `\%12\relax}%
\providecommand \@@startlink[1]{}%
\providecommand \@@endlink[0]{}%
\providecommand \url  [0]{\begingroup\@sanitize@url \@url }%
\providecommand \@url [1]{\endgroup\@href {#1}{\urlprefix }}%
\providecommand \urlprefix  [0]{URL }%
\providecommand \Eprint [0]{\href }%
\providecommand \doibase [0]{http://dx.doi.org/}%
\providecommand \selectlanguage [0]{\@gobble}%
\providecommand \bibinfo  [0]{\@secondoftwo}%
\providecommand \bibfield  [0]{\@secondoftwo}%
\providecommand \translation [1]{[#1]}%
\providecommand \BibitemOpen [0]{}%
\providecommand \bibitemStop [0]{}%
\providecommand \bibitemNoStop [0]{.\EOS\space}%
\providecommand \EOS [0]{\spacefactor3000\relax}%
\providecommand \BibitemShut  [1]{\csname bibitem#1\endcsname}%
\let\auto@bib@innerbib\@empty
\bibitem [{\citenamefont {Hetzheim}\ and\ \citenamefont
  {Keitel}(2009)}]{Hetzheim2009}%
  \BibitemOpen
  \bibfield  {author} {\bibinfo {author} {\bibfnamefont {H.~G.}\ \bibnamefont
  {Hetzheim}}\ and\ \bibinfo {author} {\bibfnamefont {C.~H.}\ \bibnamefont
  {Keitel}},\ }\bibfield  {title} {\enquote {\bibinfo {title} {Ionization
  dynamics versus laser intensity in laser-driven multiply charged ions},}\
  }\href {\doibase 10.1103/PhysRevLett.102.083003} {\bibfield  {journal}
  {\bibinfo  {journal} {Phys. Rev. Lett.}\ }\textbf {\bibinfo {volume} {102}},\
  \bibinfo {pages} {083003} (\bibinfo {year} {2009})}\BibitemShut {NoStop}%
\bibitem [{\citenamefont {Bauke}\ \emph {et~al.}(2011)\citenamefont {Bauke},
  \citenamefont {Hetzheim}, \citenamefont {Mocken}, \citenamefont {Ruf},\ and\
  \citenamefont {Keitel}}]{BaukeHetzheimetal2011}%
  \BibitemOpen
  \bibfield  {author} {\bibinfo {author} {\bibfnamefont {H.}~\bibnamefont
  {Bauke}}, \bibinfo {author} {\bibfnamefont {H.~G.}\ \bibnamefont {Hetzheim}},
  \bibinfo {author} {\bibfnamefont {G.~R.}\ \bibnamefont {Mocken}}, \bibinfo
  {author} {\bibfnamefont {M.}~\bibnamefont {Ruf}}, \ and\ \bibinfo {author}
  {\bibfnamefont {C.~H.}\ \bibnamefont {Keitel}},\ }\bibfield  {title}
  {\enquote {\bibinfo {title} {Relativistic ionization characteristics of
  laser-driven hydrogenlike ions},}\ }\href {\doibase
  10.1103/PhysRevA.83.063414} {\bibfield  {journal} {\bibinfo  {journal} {Phys.
  Rev. A}\ }\textbf {\bibinfo {volume} {83}},\ \bibinfo {pages} {063414}
  (\bibinfo {year} {2011})}\BibitemShut {NoStop}%
\bibitem [{\citenamefont {Ciappina}\ \emph {et~al.}(2019)\citenamefont
  {Ciappina}, \citenamefont {Popruzhenko}, \citenamefont {Bulanov},
  \citenamefont {Ditmire}, \citenamefont {Korn},\ and\ \citenamefont
  {Weber}}]{Ciappinaetal2019}%
  \BibitemOpen
  \bibfield  {author} {\bibinfo {author} {\bibfnamefont {M.~F.}\ \bibnamefont
  {Ciappina}}, \bibinfo {author} {\bibfnamefont {S.~V.}\ \bibnamefont
  {Popruzhenko}}, \bibinfo {author} {\bibfnamefont {S.~V.}\ \bibnamefont
  {Bulanov}}, \bibinfo {author} {\bibfnamefont {T.}~\bibnamefont {Ditmire}},
  \bibinfo {author} {\bibfnamefont {G.}~\bibnamefont {Korn}}, \ and\ \bibinfo
  {author} {\bibfnamefont {S.}~\bibnamefont {Weber}},\ }\bibfield  {title}
  {\enquote {\bibinfo {title} {Progress toward atomic diagnostics of ultrahigh
  laser intensities},}\ }\href {\doibase 10.1103/PhysRevA.99.043405} {\bibfield
   {journal} {\bibinfo  {journal} {Phys. Rev. A}\ }\textbf {\bibinfo {volume}
  {99}},\ \bibinfo {pages} {043405} (\bibinfo {year} {2019})}\BibitemShut
  {NoStop}%
\bibitem [{\citenamefont {Ciappina}\ and\ \citenamefont
  {Popruzhenko}(2020)}]{CiappinaPopruzhenko_2020}%
  \BibitemOpen
  \bibfield  {author} {\bibinfo {author} {\bibfnamefont {M.~F.}\ \bibnamefont
  {Ciappina}}\ and\ \bibinfo {author} {\bibfnamefont {S.~V.}\ \bibnamefont
  {Popruzhenko}},\ }\bibfield  {title} {\enquote {\bibinfo {title} {Diagnostics
  of ultra-intense laser pulses using tunneling ionization},}\ }\href {\doibase
  10.1088/1612-202x/ab6559} {\bibfield  {journal} {\bibinfo  {journal} {Laser
  Phys. Lett.}\ }\textbf {\bibinfo {volume} {17}},\ \bibinfo {pages} {025301}
  (\bibinfo {year} {2020})}\BibitemShut {NoStop}%
\bibitem [{\citenamefont {Ciappina}\ \emph {et~al.}(2020)\citenamefont
  {Ciappina}, \citenamefont {Bulanov}, \citenamefont {Ditmire}, \citenamefont
  {Korn},\ and\ \citenamefont {Weber}}]{CiappinaBulanovetal2020}%
  \BibitemOpen
  \bibfield  {author} {\bibinfo {author} {\bibfnamefont {M.~F.}\ \bibnamefont
  {Ciappina}}, \bibinfo {author} {\bibfnamefont {S.}~\bibnamefont {Bulanov}},
  \bibinfo {author} {\bibfnamefont {T.}~\bibnamefont {Ditmire}}, \bibinfo
  {author} {\bibfnamefont {G.}~\bibnamefont {Korn}}, \ and\ \bibinfo {author}
  {\bibfnamefont {S.}~\bibnamefont {Weber}},\ }\bibfield  {title} {\enquote
  {\bibinfo {title} {Towards laser intensity calibration using high-field
  ionization},}\ }in\ \href {\doibase 10.1007/978-3-030-47098-2_8} {\emph
  {\bibinfo {booktitle} {Progress in Ultrafast Intense Laser Science}}},\
  \bibinfo {series} {Topics in Applied Physics 136}, Vol.~\bibinfo {volume}
  {XV},\ \bibinfo {editor} {edited by\ \bibinfo {editor} {\bibfnamefont
  {K.}~\bibnamefont {Yamanouchi}}\ and\ \bibinfo {editor} {\bibfnamefont
  {D.}~\bibnamefont {Charalambidis}}}\ (\bibinfo  {publisher} {Springer
  Nature},\ \bibinfo {address} {Switzerland},\ \bibinfo {year} {2020})\
  Chap.~\bibinfo {chapter} {8}, pp.\ \bibinfo {pages} {149--176}\BibitemShut
  {NoStop}%
\bibitem [{\citenamefont {Rus}\ \emph {et~al.}(2015)\citenamefont {Rus},
  \citenamefont {Bakule}, \citenamefont {Kramer}, \citenamefont {Naylon},
  \citenamefont {Thoma}, \citenamefont {Green}, \citenamefont {Antipenkov},
  \citenamefont {Fibrich}, \citenamefont {Nov\'{a}k}, \citenamefont {Batysta},
  \citenamefont {Mazanec}, \citenamefont {Drouin}, \citenamefont {Kasl},
  \citenamefont {Ba\v{s}e}, \citenamefont {Peceli}, \citenamefont
  {Koub\'{\i}kov\'{a}}, \citenamefont {Trojek}, \citenamefont {Boge},
  \citenamefont {Lagron}, \citenamefont {Vyhl\'{\i}dka}, \citenamefont {Weiss},
  \citenamefont {Cupal}, \citenamefont {H\v{r}eb\'{\i}\v{c}ek}, \citenamefont
  {H\v{r}\'{\i}bek}, \citenamefont {\v{D}ur\'{a}k}, \citenamefont {Polan},
  \citenamefont {Ko\v{s}elja}, \citenamefont {Korn}, \citenamefont
  {Hor\'{a}\v{c}ek}, \citenamefont {Hor\'{a}\v{c}ek}, \citenamefont {Himmel},
  \citenamefont {Havl\'{\i}\v{c}ek}, \citenamefont {Honsa}, \citenamefont
  {Korou\v{s}}, \citenamefont {Laub}, \citenamefont {Haefner}, \citenamefont
  {Bayramian}, \citenamefont {Spinka}, \citenamefont {Marshall}, \citenamefont
  {Johnson}, \citenamefont {Telford}, \citenamefont {Horner}, \citenamefont
  {Deri}, \citenamefont {Metzger}, \citenamefont {Schultze}, \citenamefont
  {Mason}, \citenamefont {Ertel}, \citenamefont {Lintern}, \citenamefont
  {Greenhalgh}, \citenamefont {Edwards}, \citenamefont {Hernandez-Gomez},
  \citenamefont {Collier}, \citenamefont {Ditmire}, \citenamefont {Gaul},
  \citenamefont {Martinez}, \citenamefont {Frederickson}, \citenamefont
  {Hammond}, \citenamefont {Malato}, \citenamefont {White},\ and\ \citenamefont
  {Hou\v{z}vi\v{c}ka}}]{ELIbeamlines2015}%
  \BibitemOpen
  \bibfield  {author} {\bibinfo {author} {\bibfnamefont {B.}~\bibnamefont
  {Rus}}, \bibinfo {author} {\bibfnamefont {P.}~\bibnamefont {Bakule}},
  \bibinfo {author} {\bibfnamefont {D.}~\bibnamefont {Kramer}}, \bibinfo
  {author} {\bibfnamefont {J.}~\bibnamefont {Naylon}}, \bibinfo {author}
  {\bibfnamefont {J.}~\bibnamefont {Thoma}}, \bibinfo {author} {\bibfnamefont
  {J.~T.}\ \bibnamefont {Green}}, \bibinfo {author} {\bibfnamefont
  {R.}~\bibnamefont {Antipenkov}}, \bibinfo {author} {\bibfnamefont
  {M.}~\bibnamefont {Fibrich}}, \bibinfo {author} {\bibfnamefont
  {J.}~\bibnamefont {Nov\'{a}k}}, \bibinfo {author} {\bibfnamefont
  {F.}~\bibnamefont {Batysta}}, \bibinfo {author} {\bibfnamefont
  {T.}~\bibnamefont {Mazanec}}, \bibinfo {author} {\bibfnamefont {M.~A.}\
  \bibnamefont {Drouin}}, \bibinfo {author} {\bibfnamefont {K.}~\bibnamefont
  {Kasl}}, \bibinfo {author} {\bibfnamefont {R.}~\bibnamefont {Ba\v{s}e}},
  \bibinfo {author} {\bibfnamefont {D.}~\bibnamefont {Peceli}}, \bibinfo
  {author} {\bibfnamefont {L.}~\bibnamefont {Koub\'{\i}kov\'{a}}}, \bibinfo
  {author} {\bibfnamefont {P.}~\bibnamefont {Trojek}}, \bibinfo {author}
  {\bibfnamefont {R.}~\bibnamefont {Boge}}, \bibinfo {author} {\bibfnamefont
  {J.~C.}\ \bibnamefont {Lagron}}, \bibinfo {author} {\bibfnamefont
  {{\v{S}}.}~\bibnamefont {Vyhl\'{\i}dka}}, \bibinfo {author} {\bibfnamefont
  {J.}~\bibnamefont {Weiss}}, \bibinfo {author} {\bibfnamefont
  {J.}~\bibnamefont {Cupal}}, \bibinfo {author} {\bibfnamefont
  {J.}~\bibnamefont {H\v{r}eb\'{\i}\v{c}ek}}, \bibinfo {author} {\bibfnamefont
  {P.}~\bibnamefont {H\v{r}\'{\i}bek}}, \bibinfo {author} {\bibfnamefont
  {M.}~\bibnamefont {\v{D}ur\'{a}k}}, \bibinfo {author} {\bibfnamefont
  {J.}~\bibnamefont {Polan}}, \bibinfo {author} {\bibfnamefont
  {M.}~\bibnamefont {Ko\v{s}elja}}, \bibinfo {author} {\bibfnamefont
  {G.}~\bibnamefont {Korn}}, \bibinfo {author} {\bibfnamefont {M.}~\bibnamefont
  {Hor\'{a}\v{c}ek}}, \bibinfo {author} {\bibfnamefont {J.}~\bibnamefont
  {Hor\'{a}\v{c}ek}}, \bibinfo {author} {\bibfnamefont {B.}~\bibnamefont
  {Himmel}}, \bibinfo {author} {\bibfnamefont {T.}~\bibnamefont
  {Havl\'{\i}\v{c}ek}}, \bibinfo {author} {\bibfnamefont {A.}~\bibnamefont
  {Honsa}}, \bibinfo {author} {\bibfnamefont {P.}~\bibnamefont {Korou\v{s}}},
  \bibinfo {author} {\bibfnamefont {M.}~\bibnamefont {Laub}}, \bibinfo {author}
  {\bibfnamefont {C.}~\bibnamefont {Haefner}}, \bibinfo {author} {\bibfnamefont
  {A.}~\bibnamefont {Bayramian}}, \bibinfo {author} {\bibfnamefont
  {T.}~\bibnamefont {Spinka}}, \bibinfo {author} {\bibfnamefont
  {C.}~\bibnamefont {Marshall}}, \bibinfo {author} {\bibfnamefont
  {G.}~\bibnamefont {Johnson}}, \bibinfo {author} {\bibfnamefont
  {S.}~\bibnamefont {Telford}}, \bibinfo {author} {\bibfnamefont
  {J.}~\bibnamefont {Horner}}, \bibinfo {author} {\bibfnamefont
  {B.}~\bibnamefont {Deri}}, \bibinfo {author} {\bibfnamefont {T.}~\bibnamefont
  {Metzger}}, \bibinfo {author} {\bibfnamefont {M.}~\bibnamefont {Schultze}},
  \bibinfo {author} {\bibfnamefont {P.}~\bibnamefont {Mason}}, \bibinfo
  {author} {\bibfnamefont {K.}~\bibnamefont {Ertel}}, \bibinfo {author}
  {\bibfnamefont {A.}~\bibnamefont {Lintern}}, \bibinfo {author} {\bibfnamefont
  {J.}~\bibnamefont {Greenhalgh}}, \bibinfo {author} {\bibfnamefont
  {C.}~\bibnamefont {Edwards}}, \bibinfo {author} {\bibfnamefont
  {C.}~\bibnamefont {Hernandez-Gomez}}, \bibinfo {author} {\bibfnamefont
  {J.}~\bibnamefont {Collier}}, \bibinfo {author} {\bibfnamefont
  {T.}~\bibnamefont {Ditmire}}, \bibinfo {author} {\bibfnamefont
  {E.}~\bibnamefont {Gaul}}, \bibinfo {author} {\bibfnamefont {M.}~\bibnamefont
  {Martinez}}, \bibinfo {author} {\bibfnamefont {C.}~\bibnamefont
  {Frederickson}}, \bibinfo {author} {\bibfnamefont {D.}~\bibnamefont
  {Hammond}}, \bibinfo {author} {\bibfnamefont {C.}~\bibnamefont {Malato}},
  \bibinfo {author} {\bibfnamefont {W.}~\bibnamefont {White}}, \ and\ \bibinfo
  {author} {\bibfnamefont {J.}~\bibnamefont {Hou\v{z}vi\v{c}ka}},\ }\bibfield
  {title} {\enquote {\bibinfo {title} {{ELI-Beamlines: development of next
  generation short-pulse laser systems}},}\ }in\ \href {\doibase
  10.1117/12.2184996} {\emph {\bibinfo {booktitle} {Research Using Extreme
  Light: Entering New Frontiers with Petawatt-Class Lasers II}}},\ Vol.\
  \bibinfo {volume} {9515},\ \bibinfo {editor} {edited by\ \bibinfo {editor}
  {\bibfnamefont {G.}~\bibnamefont {Korn}}\ and\ \bibinfo {editor}
  {\bibfnamefont {L.~O.}\ \bibnamefont {Silva}}},\ \bibinfo {organization}
  {International Society for Optics and Photonics}\ (\bibinfo  {publisher}
  {SPIE},\ \bibinfo {year} {2015})\ pp.\ \bibinfo {pages} {34 --
  44}\BibitemShut {NoStop}%
\bibitem [{\citenamefont {Guo}\ \emph {et~al.}(2018)\citenamefont {Guo},
  \citenamefont {Yu}, \citenamefont {Wang}, \citenamefont {Wang}, \citenamefont
  {Liu}, \citenamefont {Gan}, \citenamefont {Li}, \citenamefont {Leng},
  \citenamefont {Liang},\ and\ \citenamefont {Li}}]{Guo2018}%
  \BibitemOpen
  \bibfield  {author} {\bibinfo {author} {\bibfnamefont {Z.}~\bibnamefont
  {Guo}}, \bibinfo {author} {\bibfnamefont {L.}~\bibnamefont {Yu}}, \bibinfo
  {author} {\bibfnamefont {J.}~\bibnamefont {Wang}}, \bibinfo {author}
  {\bibfnamefont {C.}~\bibnamefont {Wang}}, \bibinfo {author} {\bibfnamefont
  {Y.}~\bibnamefont {Liu}}, \bibinfo {author} {\bibfnamefont {Z.}~\bibnamefont
  {Gan}}, \bibinfo {author} {\bibfnamefont {W.}~\bibnamefont {Li}}, \bibinfo
  {author} {\bibfnamefont {Y.}~\bibnamefont {Leng}}, \bibinfo {author}
  {\bibfnamefont {X.}~\bibnamefont {Liang}}, \ and\ \bibinfo {author}
  {\bibfnamefont {R.}~\bibnamefont {Li}},\ }\bibfield  {title} {\enquote
  {\bibinfo {title} {Improvement of the focusing ability by double deformable
  mirrors for {10-PW-level} {Ti:} sapphire chirped pulse amplification laser
  system},}\ }\href {\doibase 10.1364/OE.26.026776} {\bibfield  {journal}
  {\bibinfo  {journal} {Opt. Express}\ }\textbf {\bibinfo {volume} {26}},\
  \bibinfo {pages} {26776} (\bibinfo {year} {2018})}\BibitemShut {NoStop}%
\bibitem [{\citenamefont {Smirnov}\ and\ \citenamefont
  {Chibisov}(1965)}]{SMIRNOVCHIBISOV1966}%
  \BibitemOpen
  \bibfield  {author} {\bibinfo {author} {\bibfnamefont {B.~M.}\ \bibnamefont
  {Smirnov}}\ and\ \bibinfo {author} {\bibfnamefont {M.~I.}\ \bibnamefont
  {Chibisov}},\ }\bibfield  {title} {\enquote {\bibinfo {title} {The breaking
  up of atomic particles by an electric field and by electron collisions},}\
  }\href {http://www.jetp.ac.ru/cgi-bin/e/index/e/22/3/p585?a=list} {\bibfield
  {journal} {\bibinfo  {journal} {Zh. Eksp. Teor. Fiz.}\ }\textbf {\bibinfo
  {volume} {49}},\ \bibinfo {pages} {841} (\bibinfo {year} {1965})},\ \bibinfo
  {note} {[Sov. Phys. JETP {\bf 22}, 585 (1966) (Engl. transl.)]}\BibitemShut
  {NoStop}%
\bibitem [{\citenamefont {Perelomov}\ \emph {et~al.}(1966)\citenamefont
  {Perelomov}, \citenamefont {Popov},\ and\ \citenamefont
  {Terentev}}]{PPT_1966}%
  \BibitemOpen
  \bibfield  {author} {\bibinfo {author} {\bibfnamefont {A.~M.}\ \bibnamefont
  {Perelomov}}, \bibinfo {author} {\bibfnamefont {V.~S.}\ \bibnamefont
  {Popov}}, \ and\ \bibinfo {author} {\bibfnamefont {M.~V.}\ \bibnamefont
  {Terentev}},\ }\bibfield  {title} {\enquote {\bibinfo {title} {Ionization of
  atoms in an alternating electric field},}\ }\href
  {http://www.jetp.ac.ru/cgi-bin/e/index/e/23/5/p924?a=list} {\bibfield
  {journal} {\bibinfo  {journal} {Zh. Eksp. Teor. Fiz.}\ }\textbf {\bibinfo
  {volume} {50}},\ \bibinfo {pages} {1393} (\bibinfo {year} {1966})},\ \bibinfo
  {note} {[Sov. Phys. JETP {\bf 23} 924 (1966) (Engl. transl.)]}\BibitemShut
  {NoStop}%
\bibitem [{\citenamefont {Popov}(2004)}]{Popov_2004}%
  \BibitemOpen
  \bibfield  {author} {\bibinfo {author} {\bibfnamefont {V.~S.}\ \bibnamefont
  {Popov}},\ }\bibfield  {title} {\enquote {\bibinfo {title} {Tunnel and
  multiphoton ionization of atoms and ions in a strong laser field ({K}eldysh
  theory)},}\ }\href {\doibase 10.1070/pu2004v047n09abeh001812} {\bibfield
  {journal} {\bibinfo  {journal} {Physics-Uspekhi}\ }\textbf {\bibinfo {volume}
  {47}},\ \bibinfo {pages} {855} (\bibinfo {year} {2004})}\BibitemShut
  {NoStop}%
\bibitem [{\citenamefont {Popruzhenko}(2014)}]{V_Popruzhenko_2014}%
  \BibitemOpen
  \bibfield  {author} {\bibinfo {author} {\bibfnamefont {S.~V.}\ \bibnamefont
  {Popruzhenko}},\ }\bibfield  {title} {\enquote {\bibinfo {title} {Keldysh
  theory of strong field ionization: history, applications, difficulties and
  perspectives},}\ }\href {\doibase 10.1088/0953-4075/47/20/204001} {\bibfield
  {journal} {\bibinfo  {journal} {J. Phys. B}\ }\textbf {\bibinfo {volume}
  {47}},\ \bibinfo {pages} {204001} (\bibinfo {year} {2014})}\BibitemShut
  {NoStop}%
\bibitem [{\citenamefont {Ammosov}\ \emph {et~al.}(1986)\citenamefont
  {Ammosov}, \citenamefont {Delone},\ and\ \citenamefont
  {Krainov}}]{Ammosov1986}%
  \BibitemOpen
  \bibfield  {author} {\bibinfo {author} {\bibfnamefont {M.~V.}\ \bibnamefont
  {Ammosov}}, \bibinfo {author} {\bibfnamefont {N.~B.}\ \bibnamefont {Delone}},
  \ and\ \bibinfo {author} {\bibfnamefont {V.~P.}\ \bibnamefont {Krainov}},\
  }\bibfield  {title} {\enquote {\bibinfo {title} {Tunnel ionization of complex
  atoms and of atomic ions in an alternating electromagnetic field},}\ }\href
  {http://www.jetp.ac.ru/cgi-bin/e/index/e/64/6/p1191?a=list} {\bibfield
  {journal} {\bibinfo  {journal} {Zh. Eksp. Teor. Fiz.}\ }\textbf {\bibinfo
  {volume} {91}},\ \bibinfo {pages} {2008} (\bibinfo {year} {1986})},\ \bibinfo
  {note} {[Sov. Phys. JETP {\bf 64} 1191 (1986) (Engl. transl.)]}\BibitemShut
  {NoStop}%
\bibitem [{\citenamefont {Tong}\ \emph {et~al.}(2002)\citenamefont {Tong},
  \citenamefont {Zhao},\ and\ \citenamefont {Lin}}]{TongZhaoLin2002}%
  \BibitemOpen
  \bibfield  {author} {\bibinfo {author} {\bibfnamefont {X.~M.}\ \bibnamefont
  {Tong}}, \bibinfo {author} {\bibfnamefont {Z.~X.}\ \bibnamefont {Zhao}}, \
  and\ \bibinfo {author} {\bibfnamefont {C.~D.}\ \bibnamefont {Lin}},\
  }\bibfield  {title} {\enquote {\bibinfo {title} {Theory of molecular
  tunneling ionization},}\ }\href {\doibase 10.1103/PhysRevA.66.033402}
  {\bibfield  {journal} {\bibinfo  {journal} {Phys. Rev. A}\ }\textbf {\bibinfo
  {volume} {66}},\ \bibinfo {pages} {033402} (\bibinfo {year}
  {2002})}\BibitemShut {NoStop}%
\bibitem [{\citenamefont {Kjeldsen}\ and\ \citenamefont
  {Madsen}(2004)}]{Kjeldsen_2004}%
  \BibitemOpen
  \bibfield  {author} {\bibinfo {author} {\bibfnamefont {T.~K.}\ \bibnamefont
  {Kjeldsen}}\ and\ \bibinfo {author} {\bibfnamefont {L.~B.}\ \bibnamefont
  {Madsen}},\ }\bibfield  {title} {\enquote {\bibinfo {title} {Strong-field
  ionization of {N$_2$}: length and velocity gauge strong-field approximation
  and tunnelling theory},}\ }\href {\doibase 10.1088/0953-4075/37/10/003}
  {\bibfield  {journal} {\bibinfo  {journal} {J. Phys. B}\ }\textbf {\bibinfo
  {volume} {37}},\ \bibinfo {pages} {2033} (\bibinfo {year}
  {2004})}\BibitemShut {NoStop}%
\bibitem [{\citenamefont {Zhao}\ \emph {et~al.}(2011)\citenamefont {Zhao},
  \citenamefont {Xu}, \citenamefont {Jin}, \citenamefont {Le},\ and\
  \citenamefont {Lin}}]{Zhao2011}%
  \BibitemOpen
  \bibfield  {author} {\bibinfo {author} {\bibfnamefont {S.-F.}\ \bibnamefont
  {Zhao}}, \bibinfo {author} {\bibfnamefont {J.}~\bibnamefont {Xu}}, \bibinfo
  {author} {\bibfnamefont {C.}~\bibnamefont {Jin}}, \bibinfo {author}
  {\bibfnamefont {A.-T.}\ \bibnamefont {Le}}, \ and\ \bibinfo {author}
  {\bibfnamefont {C.~D.}\ \bibnamefont {Lin}},\ }\bibfield  {title} {\enquote
  {\bibinfo {title} {Effect of orbital symmetry on the orientation dependence
  of strong field tunnelling ionization of nonlinear polyatomic molecules},}\
  }\href@noop {} {\bibfield  {journal} {\bibinfo  {journal} {J. Phys. B}\
  }\textbf {\bibinfo {volume} {44}},\ \bibinfo {pages} {035601} (\bibinfo
  {year} {2011})}\BibitemShut {NoStop}%
\bibitem [{\citenamefont {Kornev}\ and\ \citenamefont
  {Zon}(2015)}]{KornevZon2015}%
  \BibitemOpen
  \bibfield  {author} {\bibinfo {author} {\bibfnamefont {A.~S.}\ \bibnamefont
  {Kornev}}\ and\ \bibinfo {author} {\bibfnamefont {B.~A.}\ \bibnamefont
  {Zon}},\ }\bibfield  {title} {\enquote {\bibinfo {title} {Tunneling
  ionization of vibrationally excited nitrogen molecules},}\ }\href {\doibase
  10.1103/PhysRevA.92.033420} {\bibfield  {journal} {\bibinfo  {journal} {Phys.
  Rev. A}\ }\textbf {\bibinfo {volume} {92}},\ \bibinfo {pages} {033420}
  (\bibinfo {year} {2015})}\BibitemShut {NoStop}%
\bibitem [{\citenamefont {Kornev}\ \emph {et~al.}(2017)\citenamefont {Kornev},
  \citenamefont {Chernov},\ and\ \citenamefont {Zon}}]{KornevChernovZon2017}%
  \BibitemOpen
  \bibfield  {author} {\bibinfo {author} {\bibfnamefont {A.~S.}\ \bibnamefont
  {Kornev}}, \bibinfo {author} {\bibfnamefont {V.~E.}\ \bibnamefont {Chernov}},
  \ and\ \bibinfo {author} {\bibfnamefont {B.~A.}\ \bibnamefont {Zon}},\
  }\bibfield  {title} {\enquote {\bibinfo {title} {Laser-induced deformation of
  triatomic molecules: {I}nfluence on tunnel ionization},}\ }\href {\doibase
  10.1103/PhysRevA.96.053408} {\bibfield  {journal} {\bibinfo  {journal} {Phys.
  Rev. A}\ }\textbf {\bibinfo {volume} {96}},\ \bibinfo {pages} {053408}
  (\bibinfo {year} {2017})}\BibitemShut {NoStop}%
\bibitem [{\citenamefont {Fisher}\ \emph {et~al.}(1998)\citenamefont {Fisher},
  \citenamefont {Maron},\ and\ \citenamefont
  {Pitaevskii}}]{FisherMaronPitaevskii1998}%
  \BibitemOpen
  \bibfield  {author} {\bibinfo {author} {\bibfnamefont {D.}~\bibnamefont
  {Fisher}}, \bibinfo {author} {\bibfnamefont {Y.}~\bibnamefont {Maron}}, \
  and\ \bibinfo {author} {\bibfnamefont {L.~P.}\ \bibnamefont {Pitaevskii}},\
  }\bibfield  {title} {\enquote {\bibinfo {title} {Ionization of many-electron
  atoms by a quasistatic electric field},}\ }\href {\doibase
  10.1103/PhysRevA.58.2214} {\bibfield  {journal} {\bibinfo  {journal} {Phys.
  Rev. A}\ }\textbf {\bibinfo {volume} {58}},\ \bibinfo {pages} {2214}
  (\bibinfo {year} {1998})}\BibitemShut {NoStop}%
\bibitem [{\citenamefont {Brabec}\ \emph {et~al.}(2005)\citenamefont {Brabec},
  \citenamefont {C\^ot\'e}, \citenamefont {Boulanger},\ and\ \citenamefont
  {Ramunno}}]{Brabecetal2005}%
  \BibitemOpen
  \bibfield  {author} {\bibinfo {author} {\bibfnamefont {T.}~\bibnamefont
  {Brabec}}, \bibinfo {author} {\bibfnamefont {M.}~\bibnamefont {C\^ot\'e}},
  \bibinfo {author} {\bibfnamefont {P.}~\bibnamefont {Boulanger}}, \ and\
  \bibinfo {author} {\bibfnamefont {L.}~\bibnamefont {Ramunno}},\ }\bibfield
  {title} {\enquote {\bibinfo {title} {Theory of tunnel ionization in complex
  systems},}\ }\href {\doibase 10.1103/PhysRevLett.95.073001} {\bibfield
  {journal} {\bibinfo  {journal} {Phys. Rev. Lett.}\ }\textbf {\bibinfo
  {volume} {95}},\ \bibinfo {pages} {073001} (\bibinfo {year}
  {2005})}\BibitemShut {NoStop}%
\bibitem [{\citenamefont {Tolstikhin}\ \emph {et~al.}(2011)\citenamefont
  {Tolstikhin}, \citenamefont {Morishita},\ and\ \citenamefont
  {Madsen}}]{TolstikhinMorishitaMadsen2011}%
  \BibitemOpen
  \bibfield  {author} {\bibinfo {author} {\bibfnamefont {O.~I.}\ \bibnamefont
  {Tolstikhin}}, \bibinfo {author} {\bibfnamefont {T.}~\bibnamefont
  {Morishita}}, \ and\ \bibinfo {author} {\bibfnamefont {L.~B.}\ \bibnamefont
  {Madsen}},\ }\bibfield  {title} {\enquote {\bibinfo {title} {Theory of
  tunneling ionization of molecules: Weak-field asymptotics including dipole
  effects},}\ }\href {\doibase 10.1103/PhysRevA.84.053423} {\bibfield
  {journal} {\bibinfo  {journal} {Phys. Rev. A}\ }\textbf {\bibinfo {volume}
  {84}},\ \bibinfo {pages} {053423} (\bibinfo {year} {2011})}\BibitemShut
  {NoStop}%
\bibitem [{\citenamefont {Tolstikhin}\ \emph {et~al.}(2014)\citenamefont
  {Tolstikhin}, \citenamefont {Madsen},\ and\ \citenamefont
  {Morishita}}]{TolstikhinMadsenMorishita2014}%
  \BibitemOpen
  \bibfield  {author} {\bibinfo {author} {\bibfnamefont {O.~I.}\ \bibnamefont
  {Tolstikhin}}, \bibinfo {author} {\bibfnamefont {L.~B.}\ \bibnamefont
  {Madsen}}, \ and\ \bibinfo {author} {\bibfnamefont {T.}~\bibnamefont
  {Morishita}},\ }\bibfield  {title} {\enquote {\bibinfo {title} {Weak-field
  asymptotic theory of tunneling ionization in many-electron atomic and
  molecular systems},}\ }\href {\doibase 10.1103/PhysRevA.89.013421} {\bibfield
   {journal} {\bibinfo  {journal} {Phys. Rev. A}\ }\textbf {\bibinfo {volume}
  {89}},\ \bibinfo {pages} {013421} (\bibinfo {year} {2014})}\BibitemShut
  {NoStop}%
\bibitem [{\citenamefont {Tolstikhina}\ \emph {et~al.}(2014)\citenamefont
  {Tolstikhina}, \citenamefont {Morishita},\ and\ \citenamefont
  {Tolstikhin}}]{TolstikhinaMorishitaTolstikhin2014}%
  \BibitemOpen
  \bibfield  {author} {\bibinfo {author} {\bibfnamefont {I.~Y.}\ \bibnamefont
  {Tolstikhina}}, \bibinfo {author} {\bibfnamefont {T.}~\bibnamefont
  {Morishita}}, \ and\ \bibinfo {author} {\bibfnamefont {O.~I.}\ \bibnamefont
  {Tolstikhin}},\ }\bibfield  {title} {\enquote {\bibinfo {title} {Application
  of the many-electron weak-field asymptotic theory of tunneling ionization to
  atoms},}\ }\href {\doibase 10.1103/PhysRevA.90.053413} {\bibfield  {journal}
  {\bibinfo  {journal} {Phys. Rev. A}\ }\textbf {\bibinfo {volume} {90}},\
  \bibinfo {pages} {053413} (\bibinfo {year} {2014})}\BibitemShut {NoStop}%
\bibitem [{\citenamefont {Trinh}\ \emph {et~al.}(2015)\citenamefont {Trinh},
  \citenamefont {Tolstikhin},\ and\ \citenamefont {Morishita}}]{Trinh_2015}%
  \BibitemOpen
  \bibfield  {author} {\bibinfo {author} {\bibfnamefont {V.~H.}\ \bibnamefont
  {Trinh}}, \bibinfo {author} {\bibfnamefont {O.~I.}\ \bibnamefont
  {Tolstikhin}}, \ and\ \bibinfo {author} {\bibfnamefont {T.}~\bibnamefont
  {Morishita}},\ }\bibfield  {title} {\enquote {\bibinfo {title} {Weak-field
  asymptotic theory of tunneling ionization: benchmark analytical results for
  two-electron atoms},}\ }\href {\doibase 10.1088/0953-4075/48/6/061003}
  {\bibfield  {journal} {\bibinfo  {journal} {J. Phys. B}\ }\textbf {\bibinfo
  {volume} {48}},\ \bibinfo {pages} {061003} (\bibinfo {year}
  {2015})}\BibitemShut {NoStop}%
\bibitem [{\citenamefont {Trinh}\ \emph {et~al.}(2016)\citenamefont {Trinh},
  \citenamefont {Tolstikhin},\ and\ \citenamefont {Morishita}}]{Trinh_2016}%
  \BibitemOpen
  \bibfield  {author} {\bibinfo {author} {\bibfnamefont {V.~H.}\ \bibnamefont
  {Trinh}}, \bibinfo {author} {\bibfnamefont {O.~I.}\ \bibnamefont
  {Tolstikhin}}, \ and\ \bibinfo {author} {\bibfnamefont {T.}~\bibnamefont
  {Morishita}},\ }\bibfield  {title} {\enquote {\bibinfo {title} {First-order
  correction terms in the weak-field asymptotic theory of tunneling ionization
  in many-electron systems},}\ }\href {\doibase 10.1088/0953-4075/49/19/195603}
  {\bibfield  {journal} {\bibinfo  {journal} {J. Phys. B}\ }\textbf {\bibinfo
  {volume} {49}},\ \bibinfo {pages} {195603} (\bibinfo {year}
  {2016})}\BibitemShut {NoStop}%
\bibitem [{\citenamefont {Yue}\ \emph {et~al.}(2017)\citenamefont {Yue},
  \citenamefont {Bauch},\ and\ \citenamefont {Madsen}}]{YueBauchMadsen2017}%
  \BibitemOpen
  \bibfield  {author} {\bibinfo {author} {\bibfnamefont {L.}~\bibnamefont
  {Yue}}, \bibinfo {author} {\bibfnamefont {S.}~\bibnamefont {Bauch}}, \ and\
  \bibinfo {author} {\bibfnamefont {L.~B.}\ \bibnamefont {Madsen}},\ }\bibfield
   {title} {\enquote {\bibinfo {title} {Electron correlation in tunneling
  ionization of diatomic molecules: {A}n application of the many-electron
  weak-field asymptotic theory with a generalized-active-space partition
  scheme},}\ }\href {\doibase 10.1103/PhysRevA.96.043408} {\bibfield  {journal}
  {\bibinfo  {journal} {Phys. Rev. A}\ }\textbf {\bibinfo {volume} {96}},\
  \bibinfo {pages} {043408} (\bibinfo {year} {2017})}\BibitemShut {NoStop}%
\bibitem [{\citenamefont {Dnestryan}\ \emph {et~al.}(2019)\citenamefont
  {Dnestryan}, \citenamefont {Tolstikhin}, \citenamefont {Jensen},\ and\
  \citenamefont {Madsen}}]{DnestryanTolstikhinJensenMadsen2019}%
  \BibitemOpen
  \bibfield  {author} {\bibinfo {author} {\bibfnamefont {A.~I.}\ \bibnamefont
  {Dnestryan}}, \bibinfo {author} {\bibfnamefont {O.~I.}\ \bibnamefont
  {Tolstikhin}}, \bibinfo {author} {\bibfnamefont {F.}~\bibnamefont {Jensen}},
  \ and\ \bibinfo {author} {\bibfnamefont {L.~B.}\ \bibnamefont {Madsen}},\
  }\bibfield  {title} {\enquote {\bibinfo {title} {Torsional effects in
  strong-field ionization of molecules},}\ }\href {\doibase
  10.1103/PhysRevResearch.1.023018} {\bibfield  {journal} {\bibinfo  {journal}
  {Phys. Rev. Research}\ }\textbf {\bibinfo {volume} {1}},\ \bibinfo {pages}
  {023018} (\bibinfo {year} {2019})}\BibitemShut {NoStop}%
\bibitem [{\citenamefont {Nicolaides}\ and\ \citenamefont
  {Themelis}(1992)}]{NicolaidesThemelis1992}%
  \BibitemOpen
  \bibfield  {author} {\bibinfo {author} {\bibfnamefont {C.~A.}\ \bibnamefont
  {Nicolaides}}\ and\ \bibinfo {author} {\bibfnamefont {S.~I.}\ \bibnamefont
  {Themelis}},\ }\bibfield  {title} {\enquote {\bibinfo {title} {Theory of the
  resonances of the {LoSurdo-Stark} effect},}\ }\href {\doibase
  10.1103/PhysRevA.45.349} {\bibfield  {journal} {\bibinfo  {journal} {Phys.
  Rev. A}\ }\textbf {\bibinfo {volume} {45}},\ \bibinfo {pages} {349} (\bibinfo
  {year} {1992})}\BibitemShut {NoStop}%
\bibitem [{\citenamefont {Karlsson}\ and\ \citenamefont
  {Goscinski}(1992)}]{Karlsson_1992}%
  \BibitemOpen
  \bibfield  {author} {\bibinfo {author} {\bibfnamefont {H.~O.}\ \bibnamefont
  {Karlsson}}\ and\ \bibinfo {author} {\bibfnamefont {O.}~\bibnamefont
  {Goscinski}},\ }\bibfield  {title} {\enquote {\bibinfo {title} {Perturbed
  hydrogenic manifolds studied by the recursive residue generation method},}\
  }\href {\doibase 10.1088/0953-4075/25/23/007} {\bibfield  {journal} {\bibinfo
   {journal} {J. Phys. B}\ }\textbf {\bibinfo {volume} {25}},\ \bibinfo {pages}
  {5015} (\bibinfo {year} {1992})}\BibitemShut {NoStop}%
\bibitem [{\citenamefont {Trinh}\ \emph {et~al.}(2013)\citenamefont {Trinh},
  \citenamefont {Tolstikhin}, \citenamefont {Madsen},\ and\ \citenamefont
  {Morishita}}]{TrinhTolstikhinMadsenMorishita2013}%
  \BibitemOpen
  \bibfield  {author} {\bibinfo {author} {\bibfnamefont {V.~H.}\ \bibnamefont
  {Trinh}}, \bibinfo {author} {\bibfnamefont {O.~I.}\ \bibnamefont
  {Tolstikhin}}, \bibinfo {author} {\bibfnamefont {L.~B.}\ \bibnamefont
  {Madsen}}, \ and\ \bibinfo {author} {\bibfnamefont {T.}~\bibnamefont
  {Morishita}},\ }\bibfield  {title} {\enquote {\bibinfo {title} {First-order
  correction terms in the weak-field asymptotic theory of tunneling
  ionization},}\ }\href {\doibase 10.1103/PhysRevA.87.043426} {\bibfield
  {journal} {\bibinfo  {journal} {Phys. Rev. A}\ }\textbf {\bibinfo {volume}
  {87}},\ \bibinfo {pages} {043426} (\bibinfo {year} {2013})}\BibitemShut
  {NoStop}%
\bibitem [{\citenamefont {Nicolaides}\ and\ \citenamefont
  {Themelis}(1993)}]{NicolaidesThemelis1993}%
  \BibitemOpen
  \bibfield  {author} {\bibinfo {author} {\bibfnamefont {C.~A.}\ \bibnamefont
  {Nicolaides}}\ and\ \bibinfo {author} {\bibfnamefont {S.~I.}\ \bibnamefont
  {Themelis}},\ }\bibfield  {title} {\enquote {\bibinfo {title} {Theory and
  computation of electric-field-induced tunneling rates of polyelectronic
  atomic states},}\ }\href {\doibase 10.1103/PhysRevA.47.3122} {\bibfield
  {journal} {\bibinfo  {journal} {Phys. Rev. A}\ }\textbf {\bibinfo {volume}
  {47}},\ \bibinfo {pages} {3122} (\bibinfo {year} {1993})}\BibitemShut
  {NoStop}%
\bibitem [{\citenamefont {Scrinzi}\ \emph {et~al.}(1999)\citenamefont
  {Scrinzi}, \citenamefont {Geissler},\ and\ \citenamefont
  {Brabec}}]{ScrinziGeisslerBrabec1999}%
  \BibitemOpen
  \bibfield  {author} {\bibinfo {author} {\bibfnamefont {A.}~\bibnamefont
  {Scrinzi}}, \bibinfo {author} {\bibfnamefont {M.}~\bibnamefont {Geissler}}, \
  and\ \bibinfo {author} {\bibfnamefont {T.}~\bibnamefont {Brabec}},\
  }\bibfield  {title} {\enquote {\bibinfo {title} {Ionization above the
  {C}oulomb barrier},}\ }\href {\doibase 10.1103/PhysRevLett.83.706} {\bibfield
   {journal} {\bibinfo  {journal} {Phys. Rev. Lett.}\ }\textbf {\bibinfo
  {volume} {83}},\ \bibinfo {pages} {706} (\bibinfo {year} {1999})}\BibitemShut
  {NoStop}%
\bibitem [{\citenamefont {Scrinzi}(2000)}]{Scrinzi2000}%
  \BibitemOpen
  \bibfield  {author} {\bibinfo {author} {\bibfnamefont {A.}~\bibnamefont
  {Scrinzi}},\ }\bibfield  {title} {\enquote {\bibinfo {title} {Ionization of
  multielectron atoms by strong static electric fields},}\ }\href {\doibase
  10.1103/PhysRevA.61.041402} {\bibfield  {journal} {\bibinfo  {journal} {Phys.
  Rev. A}\ }\textbf {\bibinfo {volume} {61}},\ \bibinfo {pages} {041402(R)}
  (\bibinfo {year} {2000})}\BibitemShut {NoStop}%
\bibitem [{\citenamefont {Saenz}(2000)}]{Saenz2000}%
  \BibitemOpen
  \bibfield  {author} {\bibinfo {author} {\bibfnamefont {A.}~\bibnamefont
  {Saenz}},\ }\bibfield  {title} {\enquote {\bibinfo {title} {Enhanced
  ionization of molecular hydrogen in very strong fields},}\ }\href {\doibase
  10.1103/PhysRevA.61.051402} {\bibfield  {journal} {\bibinfo  {journal} {Phys.
  Rev. A}\ }\textbf {\bibinfo {volume} {61}},\ \bibinfo {pages} {051402(R)}
  (\bibinfo {year} {2000})}\BibitemShut {NoStop}%
\bibitem [{\citenamefont {Jagau}(2016)}]{Jagau2016}%
  \BibitemOpen
  \bibfield  {author} {\bibinfo {author} {\bibfnamefont {T.-C.}\ \bibnamefont
  {Jagau}},\ }\bibfield  {title} {\enquote {\bibinfo {title} {Investigating
  tunnel and above-barrier ionization using complex-scaled coupled-cluster
  theory},}\ }\href {\doibase 10.1063/1.4967961} {\bibfield  {journal}
  {\bibinfo  {journal} {J. Chem. Phys.}\ }\textbf {\bibinfo {volume} {145}},\
  \bibinfo {pages} {204115} (\bibinfo {year} {2016})}\BibitemShut {NoStop}%
\bibitem [{\citenamefont {Majety}\ and\ \citenamefont
  {Scrinzi}(2015{\natexlab{a}})}]{Majety_2015}%
  \BibitemOpen
  \bibfield  {author} {\bibinfo {author} {\bibfnamefont {V.~P.}\ \bibnamefont
  {Majety}}\ and\ \bibinfo {author} {\bibfnamefont {A.}~\bibnamefont
  {Scrinzi}},\ }\bibfield  {title} {\enquote {\bibinfo {title} {Static field
  ionization rates for multi-electron atoms and small molecules},}\ }\href
  {\doibase 10.1088/0953-4075/48/24/245603} {\bibfield  {journal} {\bibinfo
  {journal} {J. Phys. B}\ }\textbf {\bibinfo {volume} {48}},\ \bibinfo {pages}
  {245603} (\bibinfo {year} {2015}{\natexlab{a}})}\BibitemShut {NoStop}%
\bibitem [{\citenamefont {Majety}\ and\ \citenamefont
  {Scrinzi}(2015{\natexlab{b}})}]{MajetyScrinzi2015}%
  \BibitemOpen
  \bibfield  {author} {\bibinfo {author} {\bibfnamefont {V.~P.}\ \bibnamefont
  {Majety}}\ and\ \bibinfo {author} {\bibfnamefont {A.}~\bibnamefont
  {Scrinzi}},\ }\bibfield  {title} {\enquote {\bibinfo {title} {Dynamic
  exchange in the strong field ionization of molecules},}\ }\href {\doibase
  10.1103/PhysRevLett.115.103002} {\bibfield  {journal} {\bibinfo  {journal}
  {Phys. Rev. Lett.}\ }\textbf {\bibinfo {volume} {115}},\ \bibinfo {pages}
  {103002} (\bibinfo {year} {2015}{\natexlab{b}})}\BibitemShut {NoStop}%
\bibitem [{\citenamefont {Parker}\ \emph {et~al.}(2009)\citenamefont {Parker},
  \citenamefont {Armstrong}, \citenamefont {Boca},\ and\ \citenamefont
  {Taylor}}]{ParkerArmstrongBocaTaylor2009}%
  \BibitemOpen
  \bibfield  {author} {\bibinfo {author} {\bibfnamefont {J.~S.}\ \bibnamefont
  {Parker}}, \bibinfo {author} {\bibfnamefont {G.~S.~J.}\ \bibnamefont
  {Armstrong}}, \bibinfo {author} {\bibfnamefont {M.}~\bibnamefont {Boca}}, \
  and\ \bibinfo {author} {\bibfnamefont {K.~T.}\ \bibnamefont {Taylor}},\
  }\bibfield  {title} {\enquote {\bibinfo {title} {From the {UV} to the
  static-field limit: rates and scaling laws of intense-field ionization of
  helium},}\ }\href {\doibase 10.1088/0953-4075/42/13/134011} {\bibfield
  {journal} {\bibinfo  {journal} {J. Phys. B}\ }\textbf {\bibinfo {volume}
  {42}},\ \bibinfo {pages} {134011} (\bibinfo {year} {2009})}\BibitemShut
  {NoStop}%
\bibitem [{\citenamefont {Tong}\ and\ \citenamefont
  {Lin}(2005)}]{TongLin_2005}%
  \BibitemOpen
  \bibfield  {author} {\bibinfo {author} {\bibfnamefont {X.~M.}\ \bibnamefont
  {Tong}}\ and\ \bibinfo {author} {\bibfnamefont {C.~D.}\ \bibnamefont {Lin}},\
  }\bibfield  {title} {\enquote {\bibinfo {title} {Empirical formula for static
  field ionization rates of atoms and molecules by lasers in the
  barrier-suppression regime},}\ }\href {\doibase 10.1088/0953-4075/38/15/001}
  {\bibfield  {journal} {\bibinfo  {journal} {J. Phys. B}\ }\textbf {\bibinfo
  {volume} {38}},\ \bibinfo {pages} {2593} (\bibinfo {year}
  {2005})}\BibitemShut {NoStop}%
\bibitem [{\citenamefont {Zanghellini}\ \emph {et~al.}(2003)\citenamefont
  {Zanghellini}, \citenamefont {Kitzler}, \citenamefont {Fabian}, \citenamefont
  {Brabec},\ and\ \citenamefont {Scrinzi}}]{ZanghelliniScrinzi2003}%
  \BibitemOpen
  \bibfield  {author} {\bibinfo {author} {\bibfnamefont {J.}~\bibnamefont
  {Zanghellini}}, \bibinfo {author} {\bibfnamefont {M.}~\bibnamefont
  {Kitzler}}, \bibinfo {author} {\bibfnamefont {C.}~\bibnamefont {Fabian}},
  \bibinfo {author} {\bibfnamefont {T.}~\bibnamefont {Brabec}}, \ and\ \bibinfo
  {author} {\bibfnamefont {A.}~\bibnamefont {Scrinzi}},\ }\bibfield  {title}
  {\enquote {\bibinfo {title} {An {MCTDHF} approach to multielectron dynamics
  in laser fields},}\ }\href
  {http://www.maik.ru/full/lasphys_archive/03/8/lasphys8_03p1064full.pdf}
  {\bibfield  {journal} {\bibinfo  {journal} {Laser Phys.}\ }\textbf {\bibinfo
  {volume} {13}},\ \bibinfo {pages} {1064} (\bibinfo {year}
  {2003})}\BibitemShut {NoStop}%
\bibitem [{\citenamefont {Kato}\ and\ \citenamefont
  {Kono}(2004)}]{KatoKono2004}%
  \BibitemOpen
  \bibfield  {author} {\bibinfo {author} {\bibfnamefont {T.}~\bibnamefont
  {Kato}}\ and\ \bibinfo {author} {\bibfnamefont {H.}~\bibnamefont {Kono}},\
  }\bibfield  {title} {\enquote {\bibinfo {title} {Time-dependent
  multiconfiguration theory for electronic dynamics of molecules in an intense
  laser field},}\ }\href {\doibase 10.1016/j.cplett.2004.05.106} {\bibfield
  {journal} {\bibinfo  {journal} {Chem. Phys. Lett.}\ }\textbf {\bibinfo
  {volume} {392}},\ \bibinfo {pages} {533} (\bibinfo {year}
  {2004})}\BibitemShut {NoStop}%
\bibitem [{\citenamefont {Milosevic}\ \emph {et~al.}(2002)\citenamefont
  {Milosevic}, \citenamefont {Krainov},\ and\ \citenamefont
  {Brabec}}]{MilosevicKrainovBrabec2002}%
  \BibitemOpen
  \bibfield  {author} {\bibinfo {author} {\bibfnamefont {N.}~\bibnamefont
  {Milosevic}}, \bibinfo {author} {\bibfnamefont {V.~P.}\ \bibnamefont
  {Krainov}}, \ and\ \bibinfo {author} {\bibfnamefont {T.}~\bibnamefont
  {Brabec}},\ }\bibfield  {title} {\enquote {\bibinfo {title} {Semiclassical
  {D}irac theory of tunnel ionization},}\ }\href {\doibase
  10.1103/PhysRevLett.89.193001} {\bibfield  {journal} {\bibinfo  {journal}
  {Phys. Rev. Lett.}\ }\textbf {\bibinfo {volume} {89}},\ \bibinfo {pages}
  {193001} (\bibinfo {year} {2002})}\BibitemShut {NoStop}%
\bibitem [{\citenamefont {Keldysh}(1964)}]{Keldysh1964}%
  \BibitemOpen
  \bibfield  {author} {\bibinfo {author} {\bibfnamefont {L.~V.}\ \bibnamefont
  {Keldysh}},\ }\bibfield  {title} {\enquote {\bibinfo {title} {Ionization in
  the field of a strong electromagnetic wave},}\ }\href@noop {} {\bibfield
  {journal} {\bibinfo  {journal} {Zh. \'{E}ksp. Teor. Fiz.}\ }\textbf {\bibinfo
  {volume} {47}},\ \bibinfo {pages} {1945} (\bibinfo {year} {1964})},\ \bibinfo
  {note} {[Sov. Phys. JETP \textbf{20}, 1307 (1965)]}\BibitemShut {NoStop}%
\bibitem [{\citenamefont {Scherr}\ and\ \citenamefont
  {Knight}(1963)}]{ScherrKnight1963}%
  \BibitemOpen
  \bibfield  {author} {\bibinfo {author} {\bibfnamefont {C.~W.}\ \bibnamefont
  {Scherr}}\ and\ \bibinfo {author} {\bibfnamefont {R.~E.}\ \bibnamefont
  {Knight}},\ }\bibfield  {title} {\enquote {\bibinfo {title} {Two-electron
  atoms {III.} {A} sixth-order perturbation study of the {$1^{1}S$} ground
  state},}\ }\href {\doibase 10.1103/RevModPhys.35.436} {\bibfield  {journal}
  {\bibinfo  {journal} {Rev. Mod. Phys.}\ }\textbf {\bibinfo {volume} {35}},\
  \bibinfo {pages} {436} (\bibinfo {year} {1963})}\BibitemShut {NoStop}%
\bibitem [{\citenamefont {Hochstuhl}\ \emph {et~al.}(2014)\citenamefont
  {Hochstuhl}, \citenamefont {Hinz},\ and\ \citenamefont
  {Bonitz}}]{HochstuhlHinzBonitz2014}%
  \BibitemOpen
  \bibfield  {author} {\bibinfo {author} {\bibfnamefont {D.}~\bibnamefont
  {Hochstuhl}}, \bibinfo {author} {\bibfnamefont {C.~M.}\ \bibnamefont {Hinz}},
  \ and\ \bibinfo {author} {\bibfnamefont {M.}~\bibnamefont {Bonitz}},\
  }\bibfield  {title} {\enquote {\bibinfo {title} {Time-dependent
  multiconfiguration methods for the numerical simulation of photoionization
  processes of many-electron atoms},}\ }\href {\doibase
  10.1140/epjst/e2014-02092-3} {\bibfield  {journal} {\bibinfo  {journal} {Eur.
  Phys. J. Spec. Top.}\ }\textbf {\bibinfo {volume} {223}},\ \bibinfo {pages}
  {177} (\bibinfo {year} {2014})}\BibitemShut {NoStop}%
\bibitem [{\citenamefont {Ishikawa}\ and\ \citenamefont
  {Sato}(2015)}]{IshikawaSato_Review2015}%
  \BibitemOpen
  \bibfield  {author} {\bibinfo {author} {\bibfnamefont {K.}~\bibnamefont
  {Ishikawa}}\ and\ \bibinfo {author} {\bibfnamefont {T.}~\bibnamefont
  {Sato}},\ }\bibfield  {title} {\enquote {\bibinfo {title} {A review on ab
  initio approaches for multielectron dynamics},}\ }\href {\doibase
  10.1109/JSTQE.2015.2438827} {\bibfield  {journal} {\bibinfo  {journal} {IEEE
  J. Selec. Topics Quantum Electron.}\ }\textbf {\bibinfo {volume} {21}},\
  \bibinfo {pages} {1} (\bibinfo {year} {2015})}\BibitemShut {NoStop}%
\bibitem [{\citenamefont {L\"otstedt}\ \emph {et~al.}(2017)\citenamefont
  {L\"otstedt}, \citenamefont {Kato},\ and\ \citenamefont
  {Yamanouchi}}]{Loetstedt2017}%
  \BibitemOpen
  \bibfield  {author} {\bibinfo {author} {\bibfnamefont {E.}~\bibnamefont
  {L\"otstedt}}, \bibinfo {author} {\bibfnamefont {T.}~\bibnamefont {Kato}}, \
  and\ \bibinfo {author} {\bibfnamefont {K.}~\bibnamefont {Yamanouchi}},\
  }\bibfield  {title} {\enquote {\bibinfo {title} {Multiconfiguration methods
  for time-dependent many-electron dynamics},}\ }in\ \href {\doibase
  10.1007/978-3-319-64840-8_2} {\emph {\bibinfo {booktitle} {Progress in
  Ultrafast Intense Laser Science XIII}}},\ \bibinfo {series} {Springer Series
  in Chemical Physics}, Vol.\ \bibinfo {volume} {116},\ \bibinfo {editor}
  {edited by\ \bibinfo {editor} {\bibfnamefont {K.}~\bibnamefont {Yamanouchi}},
  \bibinfo {editor} {\bibfnamefont {W.~T.}\ \bibnamefont {Hill~III}}, \ and\
  \bibinfo {editor} {\bibfnamefont {G.~G.}\ \bibnamefont {Paulus}}}\ (\bibinfo
  {publisher} {Springer International Publishing},\ \bibinfo {address}
  {Switzerland},\ \bibinfo {year} {2017})\ pp.\ \bibinfo {pages}
  {15--40}\BibitemShut {NoStop}%
\bibitem [{\citenamefont {Sato}\ \emph {et~al.}(2018)\citenamefont {Sato},
  \citenamefont {Orimo}, \citenamefont {Teramura}, \citenamefont {Tugs},\ and\
  \citenamefont {Ishikawa}}]{SatoOrimoetal2018}%
  \BibitemOpen
  \bibfield  {author} {\bibinfo {author} {\bibfnamefont {T.}~\bibnamefont
  {Sato}}, \bibinfo {author} {\bibfnamefont {Y.}~\bibnamefont {Orimo}},
  \bibinfo {author} {\bibfnamefont {T.}~\bibnamefont {Teramura}}, \bibinfo
  {author} {\bibfnamefont {O.}~\bibnamefont {Tugs}}, \ and\ \bibinfo {author}
  {\bibfnamefont {K.~L.}\ \bibnamefont {Ishikawa}},\ }\bibfield  {title}
  {\enquote {\bibinfo {title} {Time-dependent complete-active-space
  self-consistent-field method for ultrafast intense laser science},}\ }in\
  \href {\doibase 10.1007/978-3-030-03786-4_8} {\emph {\bibinfo {booktitle}
  {Progress in Ultrafast Intense Laser Science XIV}}},\ \bibinfo {series}
  {Springer Series in Chemical Physics}, Vol.\ \bibinfo {volume} {118},\
  \bibinfo {editor} {edited by\ \bibinfo {editor} {\bibfnamefont
  {K.}~\bibnamefont {Yamanouchi}}, \bibinfo {editor} {\bibfnamefont
  {P.}~\bibnamefont {Martin}}, \bibinfo {editor} {\bibfnamefont
  {M.}~\bibnamefont {Sentis}}, \bibinfo {editor} {\bibfnamefont
  {L.}~\bibnamefont {Ruxin}}, \ and\ \bibinfo {editor} {\bibfnamefont
  {D.}~\bibnamefont {Normand}}}\ (\bibinfo  {publisher} {Springer International
  Publishing},\ \bibinfo {year} {2018})\ pp.\ \bibinfo {pages}
  {143--171}\BibitemShut {NoStop}%
\bibitem [{\citenamefont {Hochstuhl}\ and\ \citenamefont
  {Bonitz}(2011)}]{Hochstuhl2011}%
  \BibitemOpen
  \bibfield  {author} {\bibinfo {author} {\bibfnamefont {D.}~\bibnamefont
  {Hochstuhl}}\ and\ \bibinfo {author} {\bibfnamefont {M.}~\bibnamefont
  {Bonitz}},\ }\bibfield  {title} {\enquote {\bibinfo {title} {Two-photon
  ionization of helium studied with the multiconfigurational time-dependent
  {H}artree-{F}ock method},}\ }\href {\doibase 10.1063/1.3553176} {\bibfield
  {journal} {\bibinfo  {journal} {J. Chem. Phys.}\ }\textbf {\bibinfo {volume}
  {134}},\ \bibinfo {pages} {084106} (\bibinfo {year} {2011})}\BibitemShut
  {NoStop}%
\bibitem [{\citenamefont {Sato}\ \emph {et~al.}(2016)\citenamefont {Sato},
  \citenamefont {Ishikawa}, \citenamefont {B\v{r}ezinov\'a}, \citenamefont
  {Lackner}, \citenamefont {Nagele},\ and\ \citenamefont
  {Burgd\"orfer}}]{SatoIshikawaetal2016}%
  \BibitemOpen
  \bibfield  {author} {\bibinfo {author} {\bibfnamefont {T.}~\bibnamefont
  {Sato}}, \bibinfo {author} {\bibfnamefont {K.~L.}\ \bibnamefont {Ishikawa}},
  \bibinfo {author} {\bibfnamefont {I.}~\bibnamefont {B\v{r}ezinov\'a}},
  \bibinfo {author} {\bibfnamefont {F.}~\bibnamefont {Lackner}}, \bibinfo
  {author} {\bibfnamefont {S.}~\bibnamefont {Nagele}}, \ and\ \bibinfo {author}
  {\bibfnamefont {J.}~\bibnamefont {Burgd\"orfer}},\ }\bibfield  {title}
  {\enquote {\bibinfo {title} {Time-dependent complete-active-space
  self-consistent-field method for atoms: {A}pplication to high-order harmonic
  generation},}\ }\href {\doibase 10.1103/PhysRevA.94.023405} {\bibfield
  {journal} {\bibinfo  {journal} {Phys. Rev. A}\ }\textbf {\bibinfo {volume}
  {94}},\ \bibinfo {pages} {023405} (\bibinfo {year} {2016})}\BibitemShut
  {NoStop}%
\bibitem [{\citenamefont {Orimo}\ \emph {et~al.}(2018)\citenamefont {Orimo},
  \citenamefont {Sato}, \citenamefont {Scrinzi},\ and\ \citenamefont
  {Ishikawa}}]{Orimoetal2018}%
  \BibitemOpen
  \bibfield  {author} {\bibinfo {author} {\bibfnamefont {Y.}~\bibnamefont
  {Orimo}}, \bibinfo {author} {\bibfnamefont {T.}~\bibnamefont {Sato}},
  \bibinfo {author} {\bibfnamefont {A.}~\bibnamefont {Scrinzi}}, \ and\
  \bibinfo {author} {\bibfnamefont {K.~L.}\ \bibnamefont {Ishikawa}},\
  }\bibfield  {title} {\enquote {\bibinfo {title} {Implementation of the
  infinite-range exterior complex scaling to the time-dependent
  complete-active-space self-consistent-field method},}\ }\href {\doibase
  10.1103/PhysRevA.97.023423} {\bibfield  {journal} {\bibinfo  {journal} {Phys.
  Rev. A}\ }\textbf {\bibinfo {volume} {97}},\ \bibinfo {pages} {023423}
  (\bibinfo {year} {2018})}\BibitemShut {NoStop}%
\bibitem [{\citenamefont {L\"otstedt}\ \emph {et~al.}(2020)\citenamefont
  {L\"otstedt}, \citenamefont {Szidarovszky}, \citenamefont {Faisal},
  \citenamefont {Kato},\ and\ \citenamefont
  {Yamanouchi}}]{LotstedtSzidarovszkyetal2020}%
  \BibitemOpen
  \bibfield  {author} {\bibinfo {author} {\bibfnamefont {E.}~\bibnamefont
  {L\"otstedt}}, \bibinfo {author} {\bibfnamefont {T.}~\bibnamefont
  {Szidarovszky}}, \bibinfo {author} {\bibfnamefont {F.~H.~M.}\ \bibnamefont
  {Faisal}}, \bibinfo {author} {\bibfnamefont {T.}~\bibnamefont {Kato}}, \ and\
  \bibinfo {author} {\bibfnamefont {K.}~\bibnamefont {Yamanouchi}},\ }\bibfield
   {title} {\enquote {\bibinfo {title} {Excited-state populations in the
  multiconfiguration time-dependent {H}artree-{F}ock method},}\ }\href
  {\doibase 10.1088/1361-6455/ab7c3b} {\bibfield  {journal} {\bibinfo
  {journal} {J. Phys. B}\ }\textbf {\bibinfo {volume} {53}},\ \bibinfo {pages}
  {105601} (\bibinfo {year} {2020})}\BibitemShut {NoStop}%
\bibitem [{\citenamefont {Zhang}\ and\ \citenamefont
  {Lambropoulos}(1995)}]{Zhang_1995}%
  \BibitemOpen
  \bibfield  {author} {\bibinfo {author} {\bibfnamefont {J.}~\bibnamefont
  {Zhang}}\ and\ \bibinfo {author} {\bibfnamefont {P.}~\bibnamefont
  {Lambropoulos}},\ }\bibfield  {title} {\enquote {\bibinfo {title}
  {Non-perturbative time-dependent theory and {ATI} in two electron atoms},}\
  }\href {\doibase 10.1088/0953-4075/28/5/001} {\bibfield  {journal} {\bibinfo
  {journal} {J. Phys. B}\ }\textbf {\bibinfo {volume} {28}},\ \bibinfo {pages}
  {L101} (\bibinfo {year} {1995})}\BibitemShut {NoStop}%
\bibitem [{\citenamefont {Smyth}\ \emph {et~al.}(1998)\citenamefont {Smyth},
  \citenamefont {Parker},\ and\ \citenamefont {Taylor}}]{SMYTH1998}%
  \BibitemOpen
  \bibfield  {author} {\bibinfo {author} {\bibfnamefont {E.~S.}\ \bibnamefont
  {Smyth}}, \bibinfo {author} {\bibfnamefont {J.~S.}\ \bibnamefont {Parker}}, \
  and\ \bibinfo {author} {\bibfnamefont {K.}~\bibnamefont {Taylor}},\
  }\bibfield  {title} {\enquote {\bibinfo {title} {Numerical integration of the
  time-dependent schr\"odinger equation for laser-driven helium},}\ }\href
  {\doibase 10.1016/S0010-4655(98)00083-6} {\bibfield  {journal} {\bibinfo
  {journal} {Comput. Phys. Commun.}\ }\textbf {\bibinfo {volume} {114}},\
  \bibinfo {pages} {1 } (\bibinfo {year} {1998})}\BibitemShut {NoStop}%
\bibitem [{\citenamefont {Laulan}\ and\ \citenamefont
  {Bachau}(2003)}]{LaulanBachau2003}%
  \BibitemOpen
  \bibfield  {author} {\bibinfo {author} {\bibfnamefont {S.}~\bibnamefont
  {Laulan}}\ and\ \bibinfo {author} {\bibfnamefont {H.}~\bibnamefont
  {Bachau}},\ }\bibfield  {title} {\enquote {\bibinfo {title} {Correlation
  effects in two-photon single and double ionization of helium},}\ }\href
  {\doibase 10.1103/PhysRevA.68.013409} {\bibfield  {journal} {\bibinfo
  {journal} {Phys. Rev. A}\ }\textbf {\bibinfo {volume} {68}},\ \bibinfo
  {pages} {013409} (\bibinfo {year} {2003})}\BibitemShut {NoStop}%
\bibitem [{\citenamefont {L\"otstedt}\ \emph {et~al.}(2018)\citenamefont
  {L\"otstedt}, \citenamefont {Kato},\ and\ \citenamefont
  {Yamanouchi}}]{LotstedtKatoYamanouchi2017}%
  \BibitemOpen
  \bibfield  {author} {\bibinfo {author} {\bibfnamefont {E.}~\bibnamefont
  {L\"otstedt}}, \bibinfo {author} {\bibfnamefont {T.}~\bibnamefont {Kato}}, \
  and\ \bibinfo {author} {\bibfnamefont {K.}~\bibnamefont {Yamanouchi}},\
  }\bibfield  {title} {\enquote {\bibinfo {title} {Time-dependent geminal
  method applied to laser-driven beryllium},}\ }\href {\doibase
  10.1103/PhysRevA.97.013423} {\bibfield  {journal} {\bibinfo  {journal} {Phys.
  Rev. A}\ }\textbf {\bibinfo {volume} {97}},\ \bibinfo {pages} {013423}
  (\bibinfo {year} {2018})}\BibitemShut {NoStop}%
\bibitem [{\citenamefont {Reinhardt}(1982)}]{Reinhardt1982}%
  \BibitemOpen
  \bibfield  {author} {\bibinfo {author} {\bibfnamefont {W.~P.}\ \bibnamefont
  {Reinhardt}},\ }\bibfield  {title} {\enquote {\bibinfo {title} {Complex
  coordinates in the theory of atomic and molecular structure and dynamics},}\
  }\href {\doibase 10.1146/annurev.pc.33.100182.001255} {\bibfield  {journal}
  {\bibinfo  {journal} {Ann. Rev. Phys. Chem.}\ }\textbf {\bibinfo {volume}
  {33}},\ \bibinfo {pages} {223} (\bibinfo {year} {1982})}\BibitemShut
  {NoStop}%
\bibitem [{\citenamefont {Kotani}(1951)}]{Kotani1951}%
  \BibitemOpen
  \bibfield  {author} {\bibinfo {author} {\bibfnamefont {M.}~\bibnamefont
  {Kotani}},\ }\href@noop {} {\emph {\bibinfo {title} {Quantum Mechanics I}}}\
  (\bibinfo  {publisher} {Iwanami Shoten},\ \bibinfo {address} {Tokyo},\
  \bibinfo {year} {1951})\BibitemShut {NoStop}%
\bibitem [{\citenamefont {Spanner}\ and\ \citenamefont
  {Patchkovskii}(2009)}]{SpannerPatchkovskii2009}%
  \BibitemOpen
  \bibfield  {author} {\bibinfo {author} {\bibfnamefont {M.}~\bibnamefont
  {Spanner}}\ and\ \bibinfo {author} {\bibfnamefont {S.}~\bibnamefont
  {Patchkovskii}},\ }\bibfield  {title} {\enquote {\bibinfo {title}
  {One-electron ionization of multielectron systems in strong nonresonant laser
  fields},}\ }\href {\doibase 10.1103/PhysRevA.80.063411} {\bibfield  {journal}
  {\bibinfo  {journal} {Phys. Rev. A}\ }\textbf {\bibinfo {volume} {80}},\
  \bibinfo {pages} {063411} (\bibinfo {year} {2009})}\BibitemShut {NoStop}%
\bibitem [{\citenamefont {Spanner}\ and\ \citenamefont
  {Patchkovskii}(2013)}]{SPANNER201310}%
  \BibitemOpen
  \bibfield  {author} {\bibinfo {author} {\bibfnamefont {M.}~\bibnamefont
  {Spanner}}\ and\ \bibinfo {author} {\bibfnamefont {S.}~\bibnamefont
  {Patchkovskii}},\ }\bibfield  {title} {\enquote {\bibinfo {title} {Molecular
  strong field ionization and high harmonic generation: {A} selection of
  computational illustrations},}\ }\href {\doibase
  10.1016/j.chemphys.2011.12.016} {\bibfield  {journal} {\bibinfo  {journal}
  {Chem. Phys.}\ }\textbf {\bibinfo {volume} {414}},\ \bibinfo {pages} {10 }
  (\bibinfo {year} {2013})}\BibitemShut {NoStop}%
\bibitem [{\citenamefont {Majety}\ \emph {et~al.}(2015)\citenamefont {Majety},
  \citenamefont {Zielinski},\ and\ \citenamefont
  {Scrinzi}}]{MajetyZielinskiScrinzi2015}%
  \BibitemOpen
  \bibfield  {author} {\bibinfo {author} {\bibfnamefont {V.~P.}\ \bibnamefont
  {Majety}}, \bibinfo {author} {\bibfnamefont {A.}~\bibnamefont {Zielinski}}, \
  and\ \bibinfo {author} {\bibfnamefont {A.}~\bibnamefont {Scrinzi}},\
  }\bibfield  {title} {\enquote {\bibinfo {title} {Photoionization of few
  electron systems: a hybrid coupled channels approach},}\ }\href {\doibase
  10.1088/1367-2630/17/6/063002} {\bibfield  {journal} {\bibinfo  {journal}
  {New J. Phys.}\ }\textbf {\bibinfo {volume} {17}},\ \bibinfo {pages} {063002}
  (\bibinfo {year} {2015})}\BibitemShut {NoStop}%
\bibitem [{\citenamefont {Zhang}\ \emph {et~al.}(2014)\citenamefont {Zhang},
  \citenamefont {Lan},\ and\ \citenamefont {Lu}}]{ZhangLanLu2014}%
  \BibitemOpen
  \bibfield  {author} {\bibinfo {author} {\bibfnamefont {Q.}~\bibnamefont
  {Zhang}}, \bibinfo {author} {\bibfnamefont {P.}~\bibnamefont {Lan}}, \ and\
  \bibinfo {author} {\bibfnamefont {P.}~\bibnamefont {Lu}},\ }\bibfield
  {title} {\enquote {\bibinfo {title} {Empirical formula for over-barrier
  strong-field ionization},}\ }\href {\doibase 10.1103/PhysRevA.90.043410}
  {\bibfield  {journal} {\bibinfo  {journal} {Phys. Rev. A}\ }\textbf {\bibinfo
  {volume} {90}},\ \bibinfo {pages} {043410} (\bibinfo {year}
  {2014})}\BibitemShut {NoStop}%
\bibitem [{Note1()}]{Note1}%
  \BibitemOpen
  \bibinfo {note} {See Supplemental material at [link inserted by PRA] for a
  text file containing the static-field ionization rates obtained by the MCTDHF
  method for $2\le Z\le 36$}\BibitemShut {NoStop}%
\bibitem [{\citenamefont {Yang}\ \emph {et~al.}(2017)\citenamefont {Yang},
  \citenamefont {Mei}, \citenamefont {Shi},\ and\ \citenamefont
  {Qiao}}]{Yangetal2017}%
  \BibitemOpen
  \bibfield  {author} {\bibinfo {author} {\bibfnamefont {S.-J.}\ \bibnamefont
  {Yang}}, \bibinfo {author} {\bibfnamefont {X.-S.}\ \bibnamefont {Mei}},
  \bibinfo {author} {\bibfnamefont {T.-Y.}\ \bibnamefont {Shi}}, \ and\
  \bibinfo {author} {\bibfnamefont {H.-X.}\ \bibnamefont {Qiao}},\ }\bibfield
  {title} {\enquote {\bibinfo {title} {Application of the
  {H}ylleraas-{$B$}-spline basis set: {S}tatic dipole polarizabilities of
  helium},}\ }\href {\doibase 10.1103/PhysRevA.95.062505} {\bibfield  {journal}
  {\bibinfo  {journal} {Phys. Rev. A}\ }\textbf {\bibinfo {volume} {95}},\
  \bibinfo {pages} {062505} (\bibinfo {year} {2017})}\BibitemShut {NoStop}%
\bibitem [{\citenamefont {Nakashima}\ and\ \citenamefont
  {Nakatsuji}(2007)}]{NakashimaNakatsuji2007}%
  \BibitemOpen
  \bibfield  {author} {\bibinfo {author} {\bibfnamefont {H.}~\bibnamefont
  {Nakashima}}\ and\ \bibinfo {author} {\bibfnamefont {H.}~\bibnamefont
  {Nakatsuji}},\ }\bibfield  {title} {\enquote {\bibinfo {title} {Solving the
  {S}chr\"odinger equation for helium atom and its isoelectronic ions with the
  free iterative complement interaction ({ICI}) method},}\ }\href {\doibase
  10.1063/1.2801981} {\bibfield  {journal} {\bibinfo  {journal} {J. Chem.
  Phys.}\ }\textbf {\bibinfo {volume} {127}},\ \bibinfo {pages} {224104}
  (\bibinfo {year} {2007})}\BibitemShut {NoStop}%
\bibitem [{\citenamefont {Haxton}\ \emph {et~al.}(2011)\citenamefont {Haxton},
  \citenamefont {Lawler},\ and\ \citenamefont {McCurdy}}]{Haxton2011}%
  \BibitemOpen
  \bibfield  {author} {\bibinfo {author} {\bibfnamefont {D.~J.}\ \bibnamefont
  {Haxton}}, \bibinfo {author} {\bibfnamefont {K.~V.}\ \bibnamefont {Lawler}},
  \ and\ \bibinfo {author} {\bibfnamefont {C.~W.}\ \bibnamefont {McCurdy}},\
  }\bibfield  {title} {\enquote {\bibinfo {title} {Multiconfiguration
  time-dependent {H}artree-{F}ock treatment of electronic and nuclear dynamics
  in diatomic molecules},}\ }\href {\doibase 10.1103/PhysRevA.83.063416}
  {\bibfield  {journal} {\bibinfo  {journal} {Phys. Rev. A}\ }\textbf {\bibinfo
  {volume} {83}},\ \bibinfo {pages} {063416} (\bibinfo {year}
  {2011})}\BibitemShut {NoStop}%
\bibitem [{\citenamefont {Moiseyev}\ \emph {et~al.}(1978)\citenamefont
  {Moiseyev}, \citenamefont {Certain},\ and\ \citenamefont
  {Weinhold}}]{MoiseyevCertainWeinhold1978}%
  \BibitemOpen
  \bibfield  {author} {\bibinfo {author} {\bibfnamefont {N.}~\bibnamefont
  {Moiseyev}}, \bibinfo {author} {\bibfnamefont {P.}~\bibnamefont {Certain}}, \
  and\ \bibinfo {author} {\bibfnamefont {F.}~\bibnamefont {Weinhold}},\
  }\bibfield  {title} {\enquote {\bibinfo {title} {Resonance properties of
  complex-rotated hamiltonians},}\ }\href {\doibase 10.1080/00268977800102631}
  {\bibfield  {journal} {\bibinfo  {journal} {Mol. Phys.}\ }\textbf {\bibinfo
  {volume} {36}},\ \bibinfo {pages} {1613} (\bibinfo {year}
  {1978})}\BibitemShut {NoStop}%
\bibitem [{\citenamefont {Bengtsson}\ \emph {et~al.}(2008)\citenamefont
  {Bengtsson}, \citenamefont {Lindroth},\ and\ \citenamefont
  {Selst\o{}}}]{BengtssonLindrothSelsto2008}%
  \BibitemOpen
  \bibfield  {author} {\bibinfo {author} {\bibfnamefont {J.}~\bibnamefont
  {Bengtsson}}, \bibinfo {author} {\bibfnamefont {E.}~\bibnamefont {Lindroth}},
  \ and\ \bibinfo {author} {\bibfnamefont {S.}~\bibnamefont {Selst\o{}}},\
  }\bibfield  {title} {\enquote {\bibinfo {title} {Solution of the
  time-dependent {S}chr\"odinger equation using uniform complex scaling},}\
  }\href {\doibase 10.1103/PhysRevA.78.032502} {\bibfield  {journal} {\bibinfo
  {journal} {Phys. Rev. A}\ }\textbf {\bibinfo {volume} {78}},\ \bibinfo
  {pages} {032502} (\bibinfo {year} {2008})}\BibitemShut {NoStop}%
\bibitem [{\citenamefont {Jackson}(1998)}]{Jackson1998}%
  \BibitemOpen
  \bibfield  {author} {\bibinfo {author} {\bibfnamefont {J.~D.}\ \bibnamefont
  {Jackson}},\ }\href@noop {} {\emph {\bibinfo {title} {Classical
  Electrodynamics}}}\ (\bibinfo  {publisher} {John Wiley and Sons, Inc.},\
  \bibinfo {address} {Hoboken, NJ},\ \bibinfo {year} {1998})\BibitemShut
  {NoStop}%
\bibitem [{\citenamefont {Sanders}\ and\ \citenamefont
  {Scherr}(1969)}]{SandersScherr1969}%
  \BibitemOpen
  \bibfield  {author} {\bibinfo {author} {\bibfnamefont {F.~C.}\ \bibnamefont
  {Sanders}}\ and\ \bibinfo {author} {\bibfnamefont {C.~W.}\ \bibnamefont
  {Scherr}},\ }\bibfield  {title} {\enquote {\bibinfo {title} {Perturbation
  study of some excited states of two-electron atoms},}\ }\href {\doibase
  10.1103/PhysRev.181.84} {\bibfield  {journal} {\bibinfo  {journal} {Phys.
  Rev.}\ }\textbf {\bibinfo {volume} {181}},\ \bibinfo {pages} {84} (\bibinfo
  {year} {1969})}\BibitemShut {NoStop}%
\bibitem [{\citenamefont {Kramida}\ \emph {et~al.}(2019)\citenamefont
  {Kramida}, \citenamefont {{Yu.~Ralchenko}}, \citenamefont {Reader},\ and\
  \citenamefont {{and NIST ASD Team}}}]{NIST_ASD2019}%
  \BibitemOpen
  \bibfield  {author} {\bibinfo {author} {\bibfnamefont {A.}~\bibnamefont
  {Kramida}}, \bibinfo {author} {\bibnamefont {{Yu.~Ralchenko}}}, \bibinfo
  {author} {\bibfnamefont {J.}~\bibnamefont {Reader}}, \ and\ \bibinfo {author}
  {\bibnamefont {{and NIST ASD Team}}},\ }\href {https://physics.nist.gov/asd}
  {}\bibinfo {howpublished} {{NIST Atomic Spectra Database (ver. 5.7.1),
  [Online]. Available: {\tt{https://physics.nist.gov/asd}} [2019, December 26].
  National Institute of Standards and Technology, Gaithersburg, MD.}} (\bibinfo
  {year} {2019})\BibitemShut {NoStop}%
\end{thebibliography}

%

\end{document}